\documentclass[aps,prl,twocolumn,notitlepage,superscriptaddress]{revtex4-2}

\usepackage{amsmath}
\usepackage{amssymb}
\usepackage{amsfonts}
\usepackage{graphicx}
\usepackage{bm}
\usepackage{epstopdf}
\usepackage[colorlinks=true]{hyperref}
\usepackage{tensor}

\begin{document}

\title{Domain wall networks as skyrmion crystals in chiral magnets}

\author{Seungho Lee}
\affiliation{Department of Physics, Korea Advanced Institute of Science and Technology, Daejeon 34141, Republic of Korea}

\author{Toshiaki Fujimori}
\affiliation{Department of Physics $\&$ 
Research and Education Center for Natural Sciences, Keio University, 4-1-1 Hiyoshi, Kanagawa 223-8521, Japan}
\affiliation{Department of Fundamental Education, Dokkyo Medical University, 880 Kitakobayashi, Mibu, Shimotsuga, Tochigi 321-0293, Japan}

\author{Muneto Nitta}
\email{nitta@phys-h.keio.ac.jp}
\affiliation{Department of Physics $\&$ 
Research and Education Center for Natural Sciences, Keio University, 4-1-1 Hiyoshi, Kanagawa 223-8521, Japan}
\affiliation{
International Institute for Sustainability with Knotted Chiral Meta Matter (WPI-SKCM$^{\,\textit{2}}$), Hiroshima University, 1-3-2 Kagamiyama, Higashi-Hiroshima, Hiroshima 739-8511, Japan
}

\author{Se Kwon Kim}
\email{sekwonkim@kaist.ac.kr}
\affiliation{Department of Physics, Korea Advanced Institute of Science and Technology, Daejeon 34141, Republic of Korea}

\begin{abstract}
We theoretically investigate the ground states of a chiral magnet with a square anisotropy and show that it supports domain wall networks as stable ground states. A domain wall junction in the domain wall network turns out to be a skyrmion with half topological charge and, therefore, the found domain wall network has a second topological nature, a skyrmion crystal. More specifically, we present a ground-state phase diagram of the chiral magnet with varying anisotropy parameters consisting of skyrmion lattices, chiral soliton lattices, and ferromagnetic states. In the presence of the square anisotropy, the skyrmion crystal forms a domain wall network. The size of domains in the domain wall network is shown to be tunable by an external magnetic field, offering a way to realize experimentally detectable domain wall networks.
\end{abstract}

\maketitle

\begin{figure}[h]
    \includegraphics[width=0.95\columnwidth]{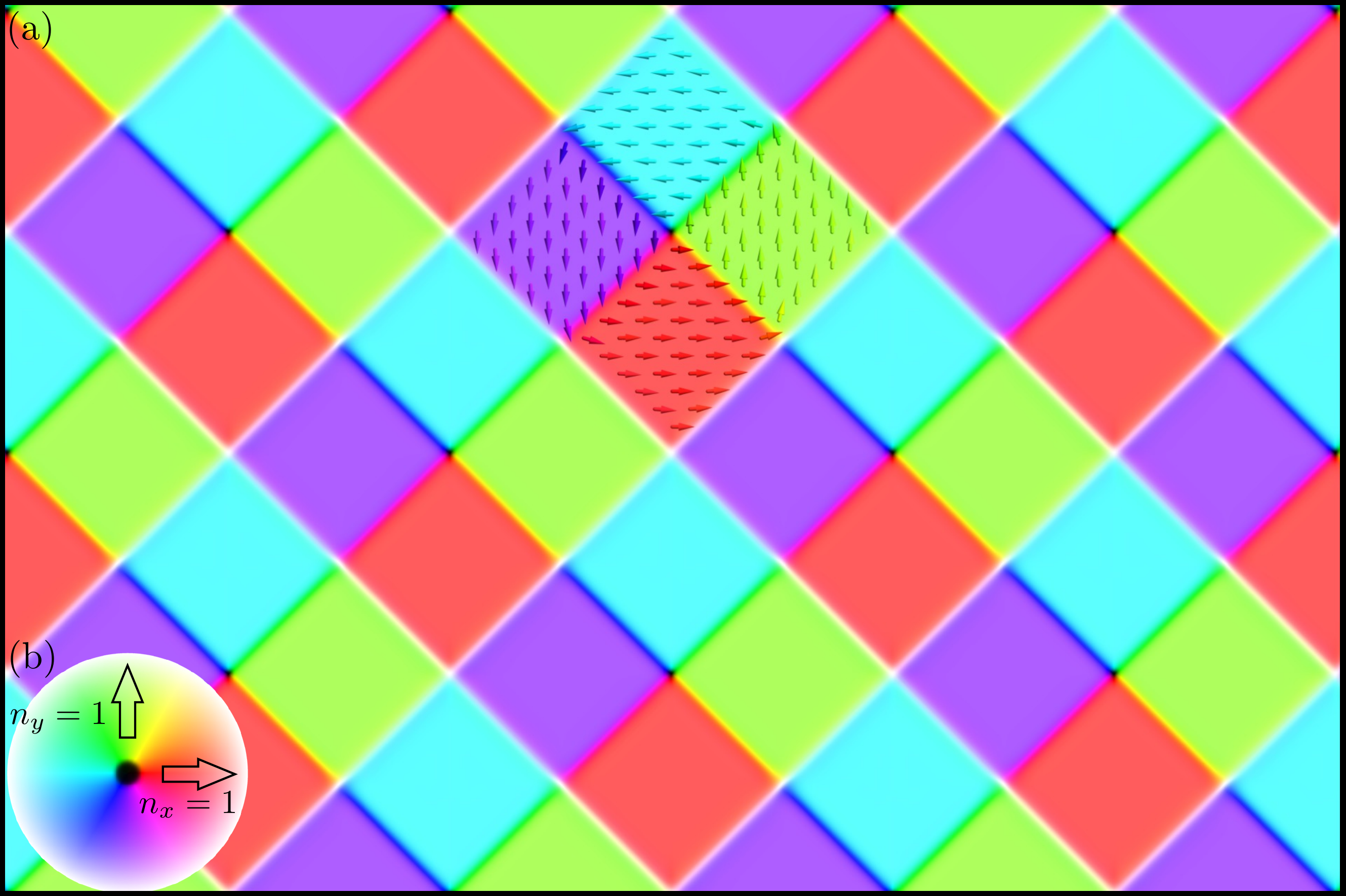}
    \caption{(a) Order-parameter configuration of the periodic domain wall network in a chiral magnet. (b) Color-coding of the arrows in the figures: Black (White) indicates the south pole $\mathbf{n}=(0,0,-1)$ (north pole $\mathbf{n}=(0,0,1)$). Hue represents the in-plane angle of the field, where red (green) corresponds to the point $\mathbf{n}=(1,0,0)$ ($\mathbf{n}=(0,1,0)$).
    }
    \label{config}
\end{figure}

\emph{Introduction.}|The field theory with spontaneously broken discrete symmetry supports topological solitons that interpolate distinct vacua, called domain walls (DWs)~\cite{Vilenkin2000, Vachaspati2006}. They can form periodic arrays known as chiral soliton lattices (CSLs)~\cite{son_prd_2008, togawa_prl_2012, eto_prd_2013, brauner_jhep_2017, amari_prb_2024, eto_prl_2025}. Such configurations have been extensively investigated theoretically, particularly in the contexts of quantum chromodynamics and chiral magnets, where CSLs can appear as ground states. The CSLs are one-dimensional periodic solutions, where the periodic modulation occurs along a single spatial direction. While the one-dimensional periodic arrays of domain walls have been thoroughly investigated, theoretical investigation of periodic domain wall networks (DWNs)~\cite{Gibbons:1999np,Carroll:1999wr,Saffin:1999au,Bazeia:1999su,Gorsky:1999hk,Oda:1999az, Nam:1999tz, Bazeia:1999xi, Eto:2005cp,Eto:2005fm,Eto:2006pg, Eto:2020vjm, Eto:2020cys, Shifman:2009zz, Kim2018c, Wang2022i, moriya_natphys_2008, heyne_prl_2010, murakami_micro_2016} in two or higher dimensions remains scarce.

When the order parameter defined in a $d$-dimensional space belongs to the $d$-dimensional unit sphere $S^d$, there exists another type of topological solitons called skyrmions characterized by the homotopy group $\pi_d(S^d)$, which were initially proposed as models for baryons in three dimensions by Tony Skyrme in the early 1960s~\cite{Skyrme1961, Skyrme1962}. 
Skyrmions in two dimensions are intensively studied in various physical systems such as nematic liquid crystals~\cite{zhao_natcom_2023, wu_lcreview_2022}, Bose-Einstein condensates~\cite{leslie_prl_2009, choi_prl_2012}, superconductors~\cite{Speight2023a, Speight2023b}, optical systems~\cite{shen_natphoto_2024}, and magnetic systems~\cite{Bogdanov:1989, Bogdanov:1995,speight_prb_2020, gobel_physreport_2021} due to their experimental realizability. In particular, magnetic skyrmions~\cite{Bogdanov:1989,Bogdanov:1995} in chiral magnets have been paid extensive attentions~\cite{Rossler:2006,doi:10.1126/science.1166767,doi:10.1038/nature09124,Han:2010by,doi:10.1038/nphys2045} due to technological utility as robust information carriers~\cite{Nagaosa2013}. Although the microscopic behaviors of all the physical systems mentioned above differ, their field theories share the same second homotopy group as the integer group, resulting in similar characteristics of topological excitations. In certain situations, skyrmions condense and thereby form a crystal, giving rise to a new topological phase of chiral magnets.

Skyrmions and DWs are two different types of topological solitons that are now closely studied together. A notable instance is a DW skyrmion that is a composite soliton of a DW and a skyrmion~\cite{Eto2005c, Nitta:2012xq, Kobayashi:2013ju, Jennings2013, Nitta:2012wi,  Nitta2013,  Gudnason2014b, Gudnason:2014hsa}. Recently, nuclear DW skyrmions were found, uncovering a new phase of nuclear matter~\cite{Eto:2023lyo,Eto:2023wul,Eto:2023tuu,Amari:2024mip}. Also, magnetic DW skyrmions have been studied in chiral magnets both theoretically~\cite{Kim:2017lsi,PhysRevB.99.184412,PhysRevB.102.094402,KBRBSK,Ross2023, Amari:2023gqv,Lee2023, Amari2024a, Gudnason:2024shv} and experimentally \cite{Nagase:2020imn,li2021magnetic,Yang:2021}. Distinct from conventional skyrmions, they can move straight along the DW without the skyrmion Hall effect, which has been hampering the practical applications of conventional skyrmions. The existing studies, however, have been restricted to the interplay of a single DW and a single skyrmion. It is an open question whether DWs and skyrmions can condense together and form a macroscopic ground state exhibiting DWNs and skyrmion crystals at the same time, which we aim to answer in this work.

The object we find here is a DWN in certain chiral magnets with the periodic junctions carrying the topological skyrmion charge defined by the second homotopy group $\pi_2 ( S^2 )$, which can thus be identified as a skyrmion crystal. Figure~\ref{config} shows the exemplary configuration of the order parameter field $\mathbf{n}:\mathbb{R}^2 \to S^2$ for a DWN solution.
Our results are envisioned to enrich the research on the interplay between DWs and skyrmions, offering a robust platform to utilize multiple solitons for technological applications.

\emph{Main results.}|In this work, we analyze the DWN as a ground state in $\mathbb{Z}_4$ chiral magnets and suggest a physical way to construct them. The $\mathbb{Z}_4$ chiral magnets, which are realizable when the system has a square atomic structure~\cite{Skomski2008}, are modeled by $O(3)$ nonlinear $\sigma$ models with the Dzyaloshinskii-Moriya (DM) term and a $\mathbb{Z}_4$ symmetric potential which contains the square anisotropy, the easy-plane anisotropy, and the Zeeman term. The DWN is a metastable or ground-state solution in this model, depending on the parameters. 
The DWN itself is a skyrmion crystal since the DW junctions carry the topological charge
\begin{equation}\label{charge}
    Q[\mathbf{n}] = \frac{1}{4\pi}\int d^2x \,   \mathbf{n}\cdot(\partial_x\mathbf{n}\times\partial_y\mathbf{n})\,.
\end{equation}
For a square-shaped DWN, there are four DW junctions in each unit cell of the network. Three DW junctions carry the topological charge of $-1/2$, and the other holds the topological charge of $1/2$. Therefore, the DWN has $Q=-1$ per unit cell.

\begin{figure}[t]
    \includegraphics[width=1\columnwidth]{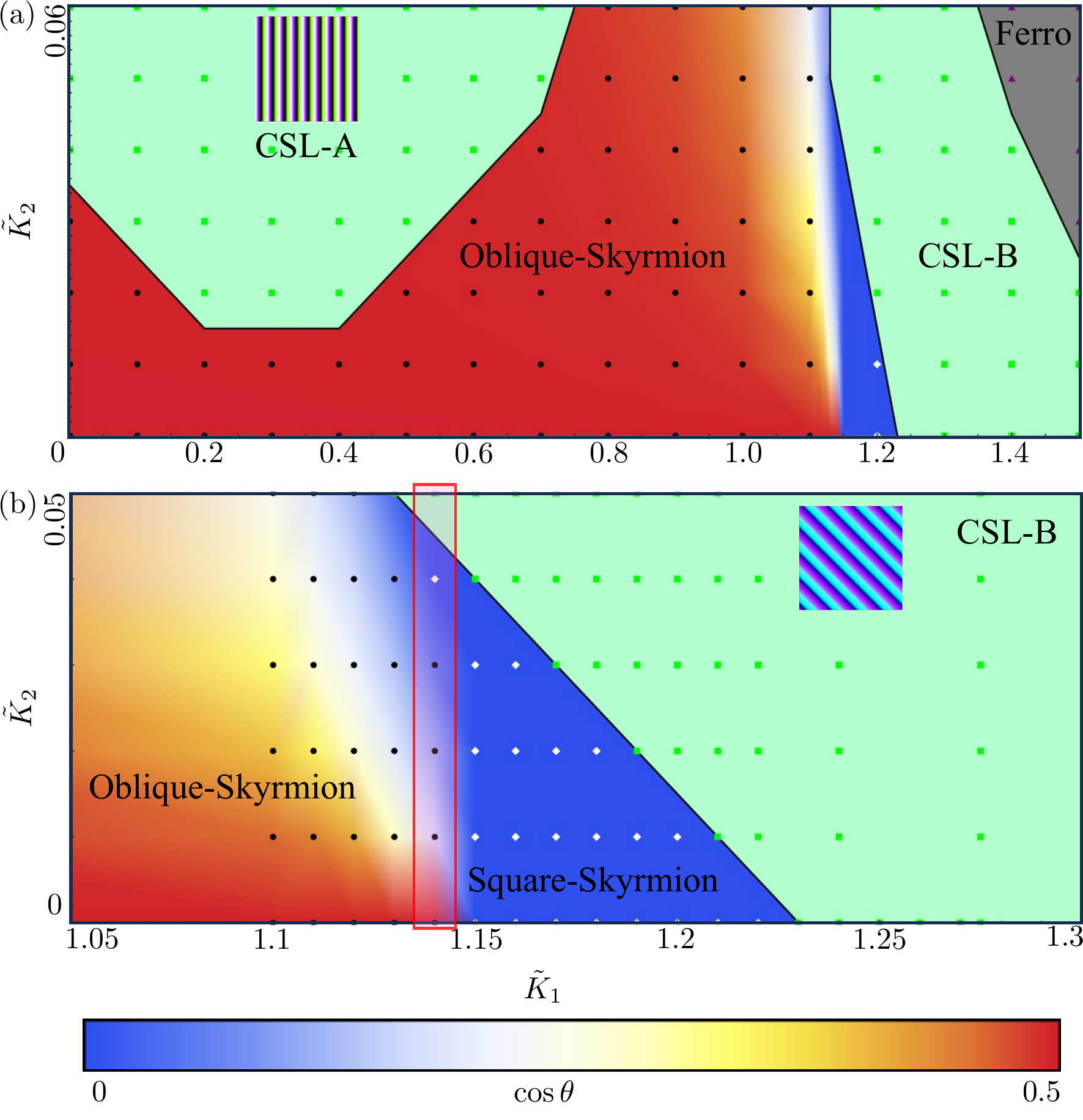}
    \caption{(a) Overall phase diagram and (b) detailed phase diagram in the vicinity of the phase boundary. Here, $\tilde{h}=0.25$. The regions and points depicted in green correspond to phases characterized by the CSL. Black and white points signify the oblique- and square-skyrmion-crystal phases, respectively. The oblique-skyrmion-crystal phase is represented through a color-coding scheme based on the angle $\theta$ between two primitive vectors of the skyrmion crystals. The gray region represents the ferromagnetic phase.}
    \label{diagram}
\end{figure}

The anisotropy dependence of ground states of $\mathbb{Z}_4$ chiral magnets with a small magnetic field is shown in Fig.~\ref{diagram}, where we find that the introduction of hitherto-neglected square anisotropy parametrized by $K_2$ in the phase diagram greatly enriches the phases of chiral magnets, constituting our first main result. Specifically, the system supports skyrmion crystals, chiral soliton lattices (CSLs), and ferromagnetic states as ground states. The ferromagnetic state is a spatially uniform state that directs one of the vacua, and CSL is a periodic solution interpolating two of four vacua, which comes in two flavors in our system. The skyrmion-crystal phase is divided into three phases: square-skyrmion-crystal phase, oblique-skyrmion-crystal phase, and hexagonal-skyrmion-crystal phase. The oblique-skyrmion phase, which continuously interpolates the hexagonal- and square-skyrmion phase, only appears in the presence of the square anisotropy. With the square anisotropy, the skyrmion crystals are also DWNs, where half-skyrmions occupy junctions where two DWs meet.

\begin{figure}[t]
    \includegraphics[width=1\columnwidth]{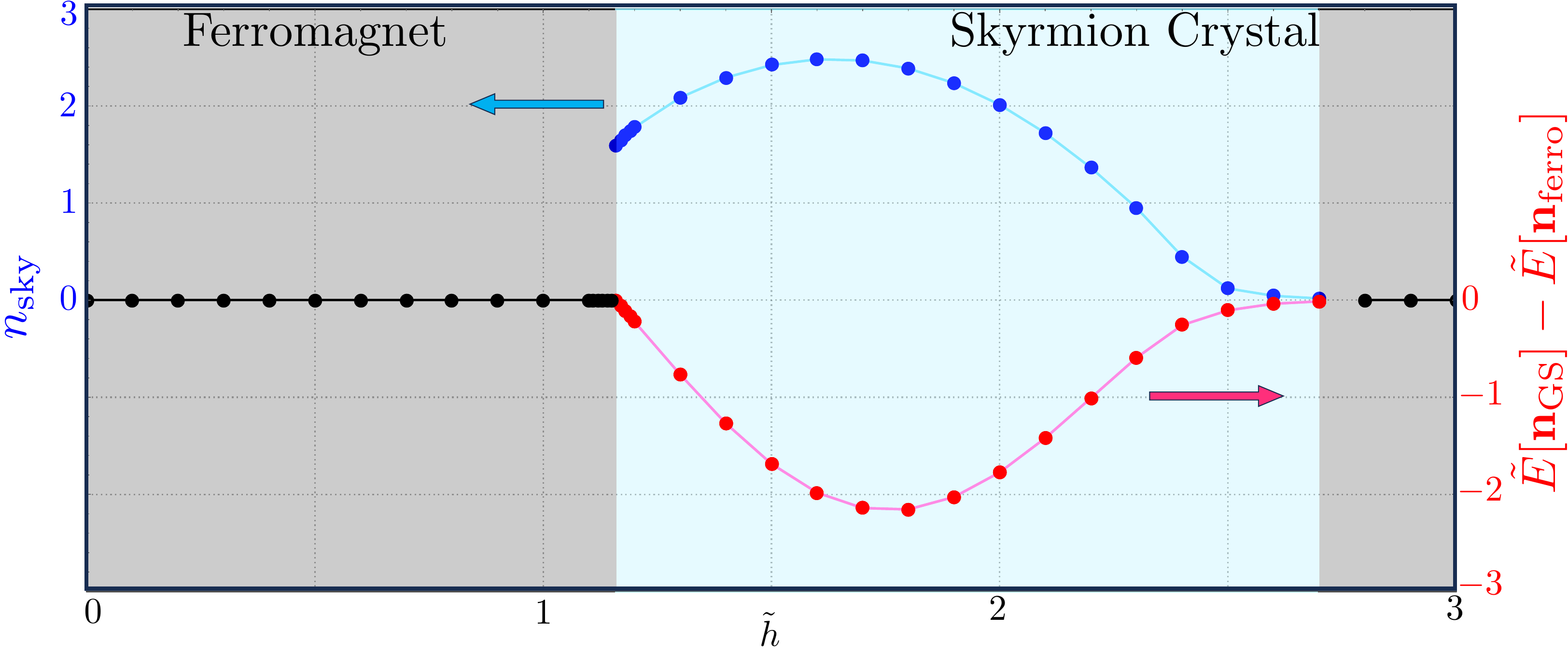}
    \caption{Skyrmion number density $n_\text{sky} = |Q|/\mathcal{A}$ and the difference of the ground-state energy density $\tilde{E}=E/\mathcal{A}$ from the ferromagnetic state. Here, $\tilde{K}_1 = 1.4$ and $\tilde{K}_2 = 0.06$.}
    \label{h_dep_plot}
\end{figure}

\begin{figure*}[t]
    \includegraphics[width=2\columnwidth]{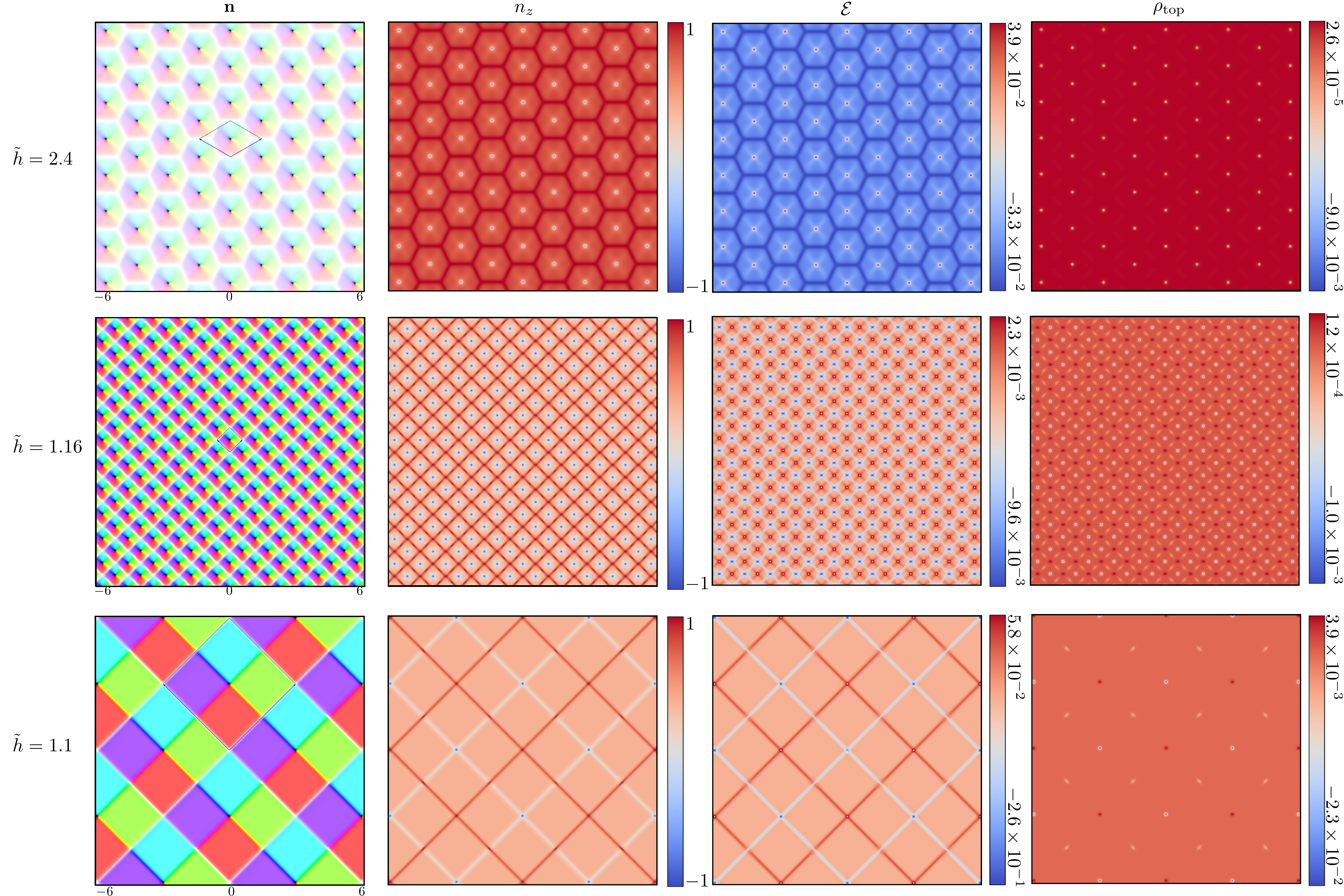}
    \caption{From left to right, field configurations, $n_z$, the energy densities $\mathcal{E}$ , and the topological charge densities $\rho_\text{top}$ of DWNs. Here, the anisotropies are fixed $\tilde{K}_1 = 1.4$ and $\tilde{K}_2 = 0.06$. From top to bottom, the magnetic field is $\tilde{h} = 2.4$, $\tilde{h} = 1.16$, and $\tilde{h}=1.1$. The scale of the plots is $12.0\times12.0$. Color-coding for the configuration is the same as in Fig.~\ref{config}. The top and the middle are ground-state solutions, while the bottom is metastable in the ferromagnetic phase. When $\tilde{h} = 2.4$, the ground state exhibits a hexagonal unit cell, whereas it exhibits a square unit cell when $\tilde{h} = 1.16$.} 
    \label{plots}
\end{figure*}

The magnetic field dependence of ground states and metastable states is also investigated, which yielded our second main results illustrated in Figs.~\ref{h_dep_plot} and~\ref{plots}. In the parameter region where the ground state is the ferromagnetic state in the absence of the magnetic field, we observe a phase transition in which the ground state changes from the DWN to the ferromagnetic state as the magnetic field decreases below a certain critical value. When we begin with the DWN ground state and decrease the magnetic field below the critical value, the DWN is shown to survive as a metastable state without decaying in the ferromagnetic phase due to its topological characteristics. We found that this metastable DWN has a larger unit-cell size compared to its ground-state counterpart, which can facilitate the experimental realization of our theoretical predictions of DWNs with the dual nature of skyrmion crystals.

\emph{Model.}|We start with an effectively two-dimensional chiral magnet whose Hamiltonian is given by
\begin{equation}\label{Hamiltonian}
    E[\mathbf{n}]=\int d^2x\left[\frac{1}{2} \sum_i |\partial_i \mathbf{n}|^2+ D\mathbf{n}\cdot\nabla\times\mathbf{n} + \mathcal{E}_0(\mathbf{n})\right]
\end{equation}
with the potential 
\begin{equation}\label{E0}
    \mathcal{E}_0(\mathbf{n})= - h n_z + K_1 n_z^2 - K_2(n_x^4+ n_y^4)\,.
\end{equation}
The first and the second terms in Eq.~\eqref{Hamiltonian} are referred to as the Dirichlet and the DM terms. The DM term allows solitonic solutions to evade Derrick's theorem~\cite{Derrick1964, Manton2009}. The first term in Eq.~\eqref{E0} is called the Zeeman term, which represents the coupling of an external magnetic field and magnets. The second and third terms in Eq.~\eqref{E0} are the easy-plane anisotropy and the square anisotropy, which is a two-dimensional analog of the cubic anisotropy~\cite{PhysRevLett.125.057201, Skomski2008, preissinger_npj_2021, wilson_prb_2014}. Note that the effect of the demagnetizing field is not considered in the model Hamiltonian~\eqref{Hamiltonian}. See the Ref.~\cite{leask_2025} for the micromagnetic analysis of the demagnetizaiton. Possible material candidates of this model~\eqref{Hamiltonian} include FeGe and Co-Zn-Mn family~\cite{yu_natmat_2011,yu_nature_2018, ukleev_prb_2024}. The potential has 4 vacua, unless the magnetic field is so strong that all vacua saturate the north or south poles. With an external magnetic field $h$ relatively small compared to the anisotropy constants, the vacua are approximately given by $(n_x, n_y, n_z)\approx(\pm 1, 0, 0)$ and $(0, \pm 1, 0)$. Hereafter we assume all constants $h,\, K_1,\, K_2,\, D$ to be positive. We use $a=1/D$ as the unit for length and dimensionless parameters $\tilde{h} = h a^2,\, \tilde{K}_1 = K_1 a^2,\, \tilde{K}_2 = K_2 a^2$, where $a$ is also the lattice constant of the corresponding discrete model~\cite{Amari2024a, Lin2015}. For numerical convenience, we set $a=1/20$.

\begin{figure}[t]
    \includegraphics[width=1\columnwidth]{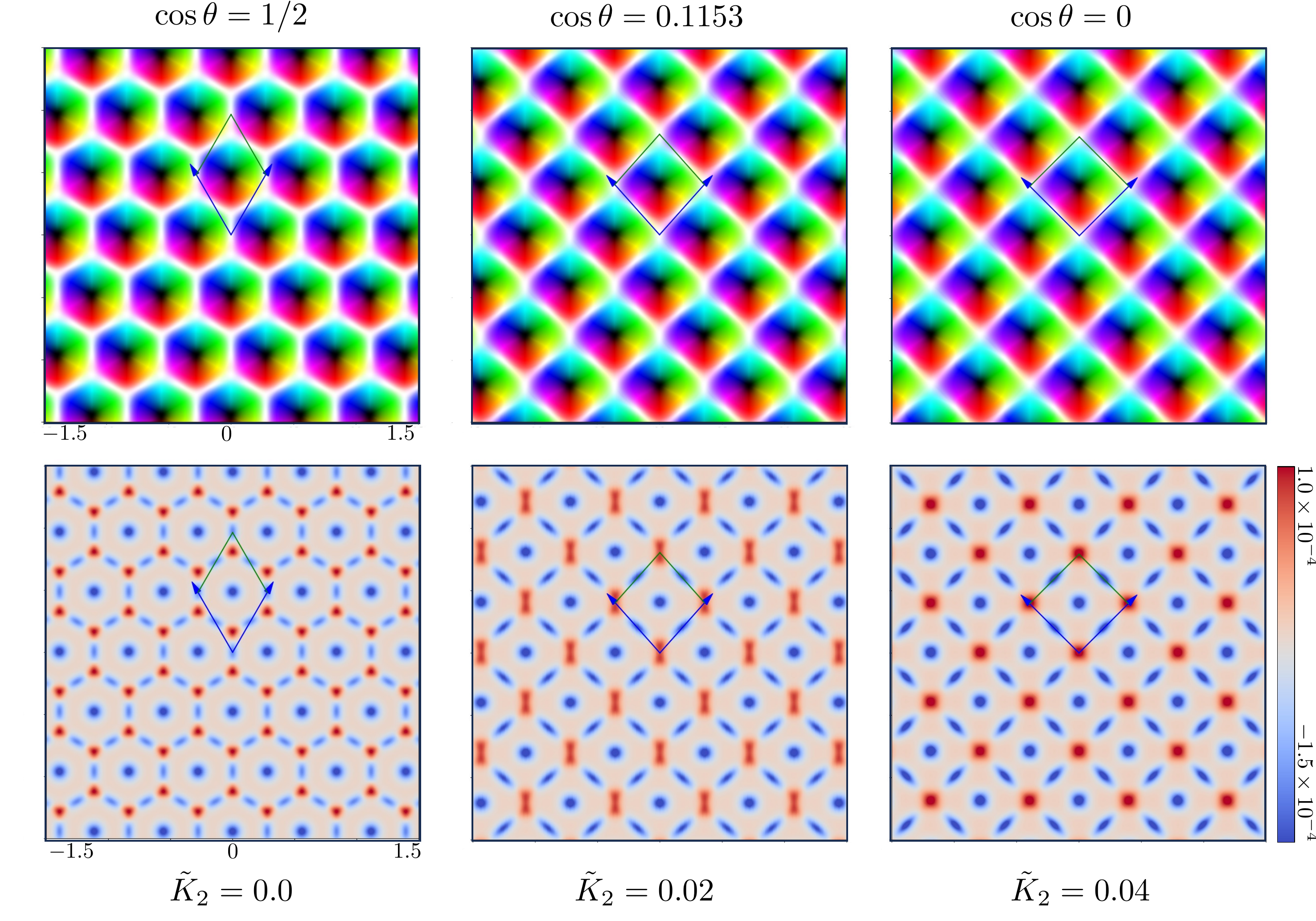}
    \caption{Phase transition induced by the square anisotropy. (Top) Order-parameter configurations and (Bottom) topological charge densities of the skyrmion crystals. Color-coding for the configuration is the same as in Fig.~\ref{config}. The blue arrows represent the primitive vectors.  Here, $\tilde{K}_1 = 1.14$ and $\tilde{h}=0.25$. The scale of the plots is $3.0\times 3.0$.}
    \label{hexoblsq}
\end{figure}

\emph{Metric optimization.}|To obtain lattice solutions of the Hamiltonian~\eqref{Hamiltonian}, we minimize the energy density on the torus for both the geometry of the unit cell and the field using the method developed by Speight {\it et.~al} in Refs.~\cite{Speight2023a, Speight2023b, Harland_jmathphys_2023, Leask2024a, Harland:2024, speight2024}, which is based on the theory of solitons on tori with nontrivial geometry~\cite{Speight2014}. This method has been successfully applied to studying soliton crystals in superconductors~\cite{Speight2023a, Speight2023b, speight2024} and nuclear matters~\cite{Harland_jmathphys_2023, Leask2024a, Harland:2024}. The scheme is as follows: First, we express the energy density as a functional of both the field and the geometry of the unit cell. Second, we minimize the functional using the arrested Newton flow~\cite{speight_prb_2020, gudnason_jhep_2020, gudnason_jhep_2022, Leask2022} with respect to the field and the geometry iteratively. When the gradients of the energy density with respect to the field and the geometry are both zero within a given tolerance, we finally obtain the lattice solution.

We define the lattice $\Lambda = \{n_1 \mathbf{v}_1 + n_2 \mathbf{v}_2 | n_i \in \mathbb{Z}, \mathbf{v}_i \in \mathbb{R}^2\}$ by two primitive vectors $\mathbf{v}_1$ and $\mathbf{v}_2$. The geometry of the unit cell is encoded by the matrix $L=(\mathbf{v}_1\,\, \mathbf{v}_2)$ and the position on the unit cell $\mathbf{x} = X_1 \mathbf{v}_1 + X_2 \mathbf{v}_2$ is parameterized by $(X_1, X_2) \in [0,1]^2$, where $X_i = (L^{-1})_{ij}x_j$. Then, the energy density functional on the torus is
\begin{equation}\label{eden}
\begin{aligned}
    \frac{E}{\mathcal{A}} =& \int_{[0,1]^2} d^2 X \bigg[\frac{1}{2}\frac{\partial \mathbf{n}}{\partial X_j} \cdot \frac{\partial \mathbf{n}}{\partial X_k} M_{ji}M_{ki}
    \\ &+ 2Dn_3 \left(\frac{\partial n_2}{\partial X_i}M_{i1} - \frac{\partial n_1}{\partial X_i}M_{i2}\right) + \mathcal{E}_0(\mathbf{n})\bigg]\,,
\end{aligned}
\end{equation}
where $M = L^{-1}$ and $\mathcal{A} = \det L$, which is the area of the unit cell. We numerically solve $\ddot{n}_\alpha = - \delta(E/\mathcal{A})/\delta n_\alpha$ and $\ddot{M}_{ij} = - \partial(E/\mathcal{A})/\partial M_{ij}$ with the arrest process that sets $\dot{\mathbf{n}}(t+\delta t)=0$ when $\ddot{\mathbf{n}}(t+\delta t)\cdot \ddot{\mathbf{n}}(t) < 0$ and sets $\dot{M}(t + \delta t) = 0$ when $\min_{ij}\ddot{M}_{ij}(t+\delta t) \ddot{M}_{ij} (t) < 0$. Using this algorithm, the system converges to one of the local minima depending on the initial conditions. We compare the energies of these local minima to determine the ground state. See the Supplemental Material for details~\footnote{The Supplemental Material includes the details of the metric-optimization method.}.

\emph{Phase diagram.}|We observe the skyrmion-crystal, CSL, and ferromagnetic phases in Fig.~\ref{diagram}. The CSL phase is further divided into type-A and type-B, defined by the wave vector of the CSL. Without the square anisotropy, the CSL solution spontaneously breaks the $U(1)$ symmetry by having the wave vector arbitrarily in the $xy$ plane. However, when $K_2 > 0$, the wave vector of the solution chooses $\hat{x}$ or $\hat{y}$ direction in the type-A phase and chooses $\hat{x}+\hat{y}$ or $\hat{x}-\hat{y}$ direction in the type-B phase. Moreover, in type-B, the CSL connects two vacua that are nearest in the target space, whereas, in type-A, the CSL connects two vacua that are next nearest in the target space, and the image on the target space makes the great circle. Since there are 4 choices for the nearest vacua and 2 choices for the next nearest vacua, the degeneracy of the type-B (A) CSL is 4 (2).

The skyrmion-crystal phase is divided into the hexagonal-, oblique-, and square-skyrmion-crystal phases defined by the angle between the primitive vectors $\mathbf{v}_1$ and $\mathbf{v}_2$. At $\tilde{K}_1 = 1.14$, which is highlighted by the red box in Fig.~\ref{diagram} (b), by tuning the value of $\tilde{K}_2$, the ground state shows four distinct phases: the hexagonal-, oblique-, square-skyrmion, and the type-B CSL. 
As shown in Fig.~\ref{hexoblsq}, by increasing the square anisotropy, the angle between two primitive vectors of the unit cell increases from $\pi/3$ to $\pi/2$. The unit cell of the skyrmion crystal has the topological charge $Q=-1$.

\emph{Field dependence.}|Since the network structure of the domains stems from the $\mathbb{Z}_4$ symmetric anisotropy, large anisotropies are favorable for observing DWNs with large domains. However, if anisotropies are too large compared to the Zeeman term, the DWN solution becomes a metastable solution, not a ground-state solution, hampering the experimental realization of DWNs. We resolve this problem by tuning the magnetic field and propose an experimental way to construct the network structure of large domains.

To this end, we investigate the magnetic-field dependence of ground states.
When $\tilde{K}_1 = 1.4$ and $\tilde{K}_2 = 0.06$, the ground state is the ferromagnetic state in the region $\tilde{h} < \tilde{h}_{c1}\approx 1.15$ and $\tilde{h} > \tilde{h}_{c2}\approx 2.7$ and the DWN in the intermediate region $\tilde{h}_{c1}<\tilde{h}<\tilde{h}_{c2}$. With a magnetic field smaller than the lower critical field $\tilde{h}_{c1}$, the ground state spontaneously breaks the $\mathbb{Z}_4$ symmetry by choosing one of four vacua, while with a magnetic field larger than the upper critical field $\tilde{h}_{c2}$, the ground state is the fully polarized state $\mathbf{n} = (0,0,1)$, which does not break the $\mathbb{Z}_4$ symmetry. 
Depending on the magnetic field, the size of the DWN varies. Figure~\ref{h_dep_plot} shows the skyrmion number density $n_\text{sky} = |Q|/\mathcal{A}$ of ground states, 
and the energy difference of the ground state from the ferromagnetic state, respectively. Note that $n_\text{sky} = 0$ for the ferromagnetic state and $n_\text{sky} = 1/\mathcal{A}$ for the DWN. 
The solution has a large unit cell as the system approaches the ferromagnetic phase. In the ferromagnetic phase $\tilde{h}<\tilde{h}_{c1}$, the DWN is still a local energy minimum, and each domain corresponds to one of four degenerate ferromagnetic states. Since spatially uniform configurations are energetically preferred, the DWN in the ferromagnetic phase has significantly large domains. 
In Fig.~\ref{plots}, we plot the configurations ${\bf n}$, $n_z$, the energy densities $\mathcal{E}$
[the integrand of Eq.~\eqref{Hamiltonian}], 
 and the topological charge densities $\rho_\text{top}$ 
 [the integrand of Eq.~\eqref{charge}] 
of DWN solutions.
The bottom plots of Fig.~\ref{plots} show the properties of the metastable DWN with $\tilde{h}=1.1$. The order parameter field is locally uniform except for the DWs and junctions. Since this solution carries the topological charge, it cannot be smoothly deformed to the topologically trivial state. Thus, one can experimentally construct the large DWN by decreasing the magnetic field starting from a ground state in the region $\tilde{h}_{c1} < \tilde{h} < \tilde{h}_{c2}$.

\emph{Conclusion.}|We have shown that a chiral magnet with a square anisotropy supports a DWN as a ground state, which is simultaneously a skyrmion crystal since the DW junction carries the topological charge. We have discussed how the square anisotropy's symmetry breaking affects properties of ground states such as the ferromagnetic states, CSLs, and skyrmion crystals. Without the square anisotropy, the CSL spontaneously breaks the $U(1)$ symmetry, although, with the square anisotropy, the CSL breaks the $\mathbb{Z}_2$ symmetry in the type-A phase and the $\mathbb{Z}_4$ symmetry in the type-B phase. The square anisotropy induces the phase transition from the hexagonal-skyrmion phase to the square-skyrmion phase through the oblique-skyrmion phase. We have also investigated the magnetic field dependence of the DWN and found magnetic-field-induced phase transitions. The size of the domains can be controlled by tuning the magnetic field, which tends to increase as the system evolves deep into the ferromagnetic phase. Due to the topological charge conservation, the DWN does not decay into the ferromagnetic state and remains as a metastable state, which can be exploited to realize our theoretical prediction of DWNs with a second topological nature as skyrmion crystals in $\mathbb{Z}_4$ chiral magnets.
Our work can be extended to DWNs and skyrmions in three-dimensional chiral magnets, where skyrmions form strings and DWs exist as surfaces glued along the skyrmion strings. The DM interaction can be interpreted as a background gauge field~\cite{schroers_scipost_2019, hill_scipost_2021, amari_jhep_2024} that stabilizes the DWNs and gives the junctions skyrmion charges. In cosmology, the effect of gauge fields on the stability of DWNs has rarely been considered. Since our work provides an example where gauge fields assist the formation of DWs, we believe that our result will stimulate further studies on DWNs endowed with gauge fields within condensed matter physics and also outside of it, such as cosmology and high-energy physics.

Furthermore, from the well-known observation that domain walls can serve as magnon waveguides~\cite{garcia-sanchez_prl_2015, lan_prx_2015, wagner_natnano_2016, sluka_natnano_2019}, DWNs are expected to act as channels for magnon transmission, and our study is anticipated to be applicable to the development of magnonic circuits.

\begin{acknowledgments} S.L. and S.K.K. were supported by Brain Pool Plus Program through the National Research Foundation of Korea funded by the Ministry of Science and ICT (NRF-2020H1D3A2A03099291), by the National Research Foundation of Korea (NRF) grant funded by the Korea government (MSIT) (NRF-2021R1C1C1006273), and by the National Research Foundation of Korea funded by the Korea Government via the SRC Center for Quantum Coherence in Condensed Matter (NRF-RS-2023-00207732). This work is also supported in part by JSPS KAKENHI [Grants No. JP21K03558 (TF), and No. JP22H01221 and No. JP23K22492 (MN)], the WPI program ``Sustainability with Knotted Chiral Meta Matter (WPI-SKCM$^2$)'' at Hiroshima University (MN).
\end{acknowledgments}

\bibliography{prb.bib}

\begin{thebibliography}{97}%
\makeatletter
\providecommand \@ifxundefined [1]{%
 \@ifx{#1\undefined}
}%
\providecommand \@ifnum [1]{%
 \ifnum #1\expandafter \@firstoftwo
 \else \expandafter \@secondoftwo
 \fi
}%
\providecommand \@ifx [1]{%
 \ifx #1\expandafter \@firstoftwo
 \else \expandafter \@secondoftwo
 \fi
}%
\providecommand \natexlab [1]{#1}%
\providecommand \enquote  [1]{``#1''}%
\providecommand \bibnamefont  [1]{#1}%
\providecommand \bibfnamefont [1]{#1}%
\providecommand \citenamefont [1]{#1}%
\providecommand \href@noop [0]{\@secondoftwo}%
\providecommand \href [0]{\begingroup \@sanitize@url \@href}%
\providecommand \@href[1]{\@@startlink{#1}\@@href}%
\providecommand \@@href[1]{\endgroup#1\@@endlink}%
\providecommand \@sanitize@url [0]{\catcode `\\12\catcode `\$12\catcode
  `\&12\catcode `\#12\catcode `\^12\catcode `\_12\catcode `\%12\relax}%
\providecommand \@@startlink[1]{}%
\providecommand \@@endlink[0]{}%
\providecommand \url  [0]{\begingroup\@sanitize@url \@url }%
\providecommand \@url [1]{\endgroup\@href {#1}{\urlprefix }}%
\providecommand \urlprefix  [0]{URL }%
\providecommand \Eprint [0]{\href }%
\providecommand \doibase [0]{https://doi.org/}%
\providecommand \selectlanguage [0]{\@gobble}%
\providecommand \bibinfo  [0]{\@secondoftwo}%
\providecommand \bibfield  [0]{\@secondoftwo}%
\providecommand \translation [1]{[#1]}%
\providecommand \BibitemOpen [0]{}%
\providecommand \bibitemStop [0]{}%
\providecommand \bibitemNoStop [0]{.\EOS\space}%
\providecommand \EOS [0]{\spacefactor3000\relax}%
\providecommand \BibitemShut  [1]{\csname bibitem#1\endcsname}%
\let\auto@bib@innerbib\@empty
\bibitem [{\citenamefont {Vilenkin}\ and\ \citenamefont
  {Shellard}(2000)}]{Vilenkin2000}%
  \BibitemOpen
  \bibfield  {author} {\bibinfo {author} {\bibfnamefont {A.}~\bibnamefont
  {Vilenkin}}\ and\ \bibinfo {author} {\bibfnamefont {E.~P.~S.}\ \bibnamefont
  {Shellard}},\ }\href@noop {} {\emph {\bibinfo {title} {Cosmic {{Strings}} and
  {{Other Topological Defects}}}}}\ (\bibinfo  {publisher} {Cambridge
  University Press},\ \bibinfo {year} {2000})\BibitemShut {NoStop}%
\bibitem [{\citenamefont {Vachaspati}(2006)}]{Vachaspati2006}%
  \BibitemOpen
  \bibfield  {author} {\bibinfo {author} {\bibfnamefont {T.}~\bibnamefont
  {Vachaspati}},\ }\href@noop {} {\emph {\bibinfo {title} {Kinks and {{Domain
  Walls}}: {{An Introduction}} to {{Classical}} and {{Quantum Solitons}}}}}\
  (\bibinfo  {publisher} {Cambridge University Press},\ \bibinfo {address}
  {Cambridge},\ \bibinfo {year} {2006})\BibitemShut {NoStop}%
\bibitem [{\citenamefont {Son}\ and\ \citenamefont
  {Stephanov}(2008)}]{son_prd_2008}%
  \BibitemOpen
  \bibfield  {author} {\bibinfo {author} {\bibfnamefont {D.~T.}\ \bibnamefont
  {Son}}\ and\ \bibinfo {author} {\bibfnamefont {M.~A.}\ \bibnamefont
  {Stephanov}},\ }\bibfield  {title} {\bibinfo {title} {Axial anomaly and
  magnetism of nuclear and quark matter},\ }\href
  {https://doi.org/10.1103/PhysRevD.77.014021} {\bibfield  {journal} {\bibinfo
  {journal} {Phys. Rev. D}\ }\textbf {\bibinfo {volume} {77}},\ \bibinfo
  {pages} {014021} (\bibinfo {year} {2008})}\BibitemShut {NoStop}%
\bibitem [{\citenamefont {Togawa}\ \emph {et~al.}(2012)\citenamefont {Togawa},
  \citenamefont {Koyama}, \citenamefont {Takayanagi}, \citenamefont {Mori},
  \citenamefont {Kousaka}, \citenamefont {Akimitsu}, \citenamefont {Nishihara},
  \citenamefont {Inoue}, \citenamefont {Ovchinnikov},\ and\ \citenamefont
  {Kishine}}]{togawa_prl_2012}%
  \BibitemOpen
  \bibfield  {author} {\bibinfo {author} {\bibfnamefont {Y.}~\bibnamefont
  {Togawa}}, \bibinfo {author} {\bibfnamefont {T.}~\bibnamefont {Koyama}},
  \bibinfo {author} {\bibfnamefont {K.}~\bibnamefont {Takayanagi}}, \bibinfo
  {author} {\bibfnamefont {S.}~\bibnamefont {Mori}}, \bibinfo {author}
  {\bibfnamefont {Y.}~\bibnamefont {Kousaka}}, \bibinfo {author} {\bibfnamefont
  {J.}~\bibnamefont {Akimitsu}}, \bibinfo {author} {\bibfnamefont
  {S.}~\bibnamefont {Nishihara}}, \bibinfo {author} {\bibfnamefont
  {K.}~\bibnamefont {Inoue}}, \bibinfo {author} {\bibfnamefont {A.~S.}\
  \bibnamefont {Ovchinnikov}},\ and\ \bibinfo {author} {\bibfnamefont
  {J.}~\bibnamefont {Kishine}},\ }\bibfield  {title} {\bibinfo {title} {Chiral
  magnetic soliton lattice on a chiral helimagnet},\ }\href
  {https://doi.org/10.1103/PhysRevLett.108.107202} {\bibfield  {journal}
  {\bibinfo  {journal} {Phys. Rev. Lett.}\ }\textbf {\bibinfo {volume} {108}},\
  \bibinfo {pages} {107202} (\bibinfo {year} {2012})}\BibitemShut {NoStop}%
\bibitem [{\citenamefont {Eto}\ \emph {et~al.}(2013)\citenamefont {Eto},
  \citenamefont {Hashimoto},\ and\ \citenamefont {Hatsuda}}]{eto_prd_2013}%
  \BibitemOpen
  \bibfield  {author} {\bibinfo {author} {\bibfnamefont {M.}~\bibnamefont
  {Eto}}, \bibinfo {author} {\bibfnamefont {K.}~\bibnamefont {Hashimoto}},\
  and\ \bibinfo {author} {\bibfnamefont {T.}~\bibnamefont {Hatsuda}},\
  }\bibfield  {title} {\bibinfo {title} {Ferromagnetic neutron stars: Axial
  anomaly, dense neutron matter, and pionic wall},\ }\href
  {https://doi.org/10.1103/PhysRevD.88.081701} {\bibfield  {journal} {\bibinfo
  {journal} {Phys. Rev. D}\ }\textbf {\bibinfo {volume} {88}},\ \bibinfo
  {pages} {081701} (\bibinfo {year} {2013})}\BibitemShut {NoStop}%
\bibitem [{\citenamefont {Brauner}\ and\ \citenamefont
  {Yamamoto}(2017)}]{brauner_jhep_2017}%
  \BibitemOpen
  \bibfield  {author} {\bibinfo {author} {\bibfnamefont {T.}~\bibnamefont
  {Brauner}}\ and\ \bibinfo {author} {\bibfnamefont {N.}~\bibnamefont
  {Yamamoto}},\ }\bibfield  {title} {\bibinfo {title} {Chiral soliton lattice
  and charged pion condensation in strong magnetic fields},\ }\href
  {https://doi.org/10.1007/JHEP04(2017)132} {\bibfield  {journal} {\bibinfo
  {journal} {Journal of High Energy Physics}\ }\textbf {\bibinfo {volume}
  {2017}},\ \bibinfo {pages} {132} (\bibinfo {year} {2017})}\BibitemShut
  {NoStop}%
\bibitem [{\citenamefont {Amari}\ \emph
  {et~al.}(2024{\natexlab{a}})\citenamefont {Amari}, \citenamefont {Ross},\
  and\ \citenamefont {Nitta}}]{amari_prb_2024}%
  \BibitemOpen
  \bibfield  {author} {\bibinfo {author} {\bibfnamefont {Y.}~\bibnamefont
  {Amari}}, \bibinfo {author} {\bibfnamefont {C.}~\bibnamefont {Ross}},\ and\
  \bibinfo {author} {\bibfnamefont {M.}~\bibnamefont {Nitta}},\ }\bibfield
  {title} {\bibinfo {title} {Domain-wall skyrmion chain and domain-wall
  bimerons in chiral magnets},\ }\href
  {https://doi.org/10.1103/PhysRevB.109.104426} {\bibfield  {journal} {\bibinfo
   {journal} {Phys. Rev. B}\ }\textbf {\bibinfo {volume} {109}},\ \bibinfo
  {pages} {104426} (\bibinfo {year} {2024}{\natexlab{a}})}\BibitemShut
  {NoStop}%
\bibitem [{\citenamefont {Eto}\ \emph {et~al.}(2025)\citenamefont {Eto},
  \citenamefont {Nishimura},\ and\ \citenamefont {Nitta}}]{eto_prl_2025}%
  \BibitemOpen
  \bibfield  {author} {\bibinfo {author} {\bibfnamefont {M.}~\bibnamefont
  {Eto}}, \bibinfo {author} {\bibfnamefont {K.}~\bibnamefont {Nishimura}},\
  and\ \bibinfo {author} {\bibfnamefont {M.}~\bibnamefont {Nitta}},\ }\bibfield
   {title} {\bibinfo {title} {Domain-wall skyrmion phase in dense qcd at strong
  magnetic fields using leading-order chiral perturbation theory},\ }\href
  {https://doi.org/10.1103/PhysRevLett.134.181902} {\bibfield  {journal}
  {\bibinfo  {journal} {Phys. Rev. Lett.}\ }\textbf {\bibinfo {volume} {134}},\
  \bibinfo {pages} {181902} (\bibinfo {year} {2025})}\BibitemShut {NoStop}%
\bibitem [{\citenamefont {Gibbons}\ and\ \citenamefont
  {Townsend}(1999)}]{Gibbons:1999np}%
  \BibitemOpen
  \bibfield  {author} {\bibinfo {author} {\bibfnamefont {G.~W.}\ \bibnamefont
  {Gibbons}}\ and\ \bibinfo {author} {\bibfnamefont {P.~K.}\ \bibnamefont
  {Townsend}},\ }\bibfield  {title} {\bibinfo {title} {{A Bogomolny equation
  for intersecting domain walls}},\ }\href
  {https://doi.org/10.1103/PhysRevLett.83.1727} {\bibfield  {journal} {\bibinfo
   {journal} {Phys. Rev. Lett.}\ }\textbf {\bibinfo {volume} {83}},\ \bibinfo
  {pages} {1727} (\bibinfo {year} {1999})},\ \Eprint
  {https://arxiv.org/abs/hep-th/9905196} {arXiv:hep-th/9905196} \BibitemShut
  {NoStop}%
\bibitem [{\citenamefont {Carroll}\ \emph {et~al.}(2000)\citenamefont
  {Carroll}, \citenamefont {Hellerman},\ and\ \citenamefont
  {Trodden}}]{Carroll:1999wr}%
  \BibitemOpen
  \bibfield  {author} {\bibinfo {author} {\bibfnamefont {S.~M.}\ \bibnamefont
  {Carroll}}, \bibinfo {author} {\bibfnamefont {S.}~\bibnamefont {Hellerman}},\
  and\ \bibinfo {author} {\bibfnamefont {M.}~\bibnamefont {Trodden}},\
  }\bibfield  {title} {\bibinfo {title} {{Domain wall junctions are 1/4 - BPS
  states}},\ }\href {https://doi.org/10.1103/PhysRevD.61.065001} {\bibfield
  {journal} {\bibinfo  {journal} {Phys. Rev. D}\ }\textbf {\bibinfo {volume}
  {61}},\ \bibinfo {pages} {065001} (\bibinfo {year} {2000})},\ \Eprint
  {https://arxiv.org/abs/hep-th/9905217} {arXiv:hep-th/9905217} \BibitemShut
  {NoStop}%
\bibitem [{\citenamefont {Saffin}(1999)}]{Saffin:1999au}%
  \BibitemOpen
  \bibfield  {author} {\bibinfo {author} {\bibfnamefont {P.~M.}\ \bibnamefont
  {Saffin}},\ }\bibfield  {title} {\bibinfo {title} {{Tiling with almost BPS
  junctions}},\ }\href {https://doi.org/10.1103/PhysRevLett.83.4249} {\bibfield
   {journal} {\bibinfo  {journal} {Phys. Rev. Lett.}\ }\textbf {\bibinfo
  {volume} {83}},\ \bibinfo {pages} {4249} (\bibinfo {year} {1999})},\ \Eprint
  {https://arxiv.org/abs/hep-th/9907066} {arXiv:hep-th/9907066} \BibitemShut
  {NoStop}%
\bibitem [{\citenamefont {Bazeia}\ and\ \citenamefont
  {Brito}(2000{\natexlab{a}})}]{Bazeia:1999su}%
  \BibitemOpen
  \bibfield  {author} {\bibinfo {author} {\bibfnamefont {D.}~\bibnamefont
  {Bazeia}}\ and\ \bibinfo {author} {\bibfnamefont {F.~A.}\ \bibnamefont
  {Brito}},\ }\bibfield  {title} {\bibinfo {title} {{Tiling the plane without
  supersymmetry}},\ }\href {https://doi.org/10.1103/PhysRevLett.84.1094}
  {\bibfield  {journal} {\bibinfo  {journal} {Phys. Rev. Lett.}\ }\textbf
  {\bibinfo {volume} {84}},\ \bibinfo {pages} {1094} (\bibinfo {year}
  {2000}{\natexlab{a}})},\ \Eprint {https://arxiv.org/abs/hep-th/9908090}
  {arXiv:hep-th/9908090} \BibitemShut {NoStop}%
\bibitem [{\citenamefont {Gorsky}\ and\ \citenamefont
  {Shifman}(2000)}]{Gorsky:1999hk}%
  \BibitemOpen
  \bibfield  {author} {\bibinfo {author} {\bibfnamefont {A.}~\bibnamefont
  {Gorsky}}\ and\ \bibinfo {author} {\bibfnamefont {M.~A.}\ \bibnamefont
  {Shifman}},\ }\bibfield  {title} {\bibinfo {title} {{More on the tensorial
  central charges in N=1 supersymmetric gauge theories (BPS wall junctions and
  strings)}},\ }\href {https://doi.org/10.1103/PhysRevD.61.085001} {\bibfield
  {journal} {\bibinfo  {journal} {Phys. Rev. D}\ }\textbf {\bibinfo {volume}
  {61}},\ \bibinfo {pages} {085001} (\bibinfo {year} {2000})},\ \Eprint
  {https://arxiv.org/abs/hep-th/9909015} {arXiv:hep-th/9909015} \BibitemShut
  {NoStop}%
\bibitem [{\citenamefont {Oda}\ \emph {et~al.}(1999)\citenamefont {Oda},
  \citenamefont {Ito}, \citenamefont {Naganuma},\ and\ \citenamefont
  {Sakai}}]{Oda:1999az}%
  \BibitemOpen
  \bibfield  {author} {\bibinfo {author} {\bibfnamefont {H.}~\bibnamefont
  {Oda}}, \bibinfo {author} {\bibfnamefont {K.}~\bibnamefont {Ito}}, \bibinfo
  {author} {\bibfnamefont {M.}~\bibnamefont {Naganuma}},\ and\ \bibinfo
  {author} {\bibfnamefont {N.}~\bibnamefont {Sakai}},\ }\bibfield  {title}
  {\bibinfo {title} {{An Exact solution of BPS domain wall junction}},\ }\href
  {https://doi.org/10.1016/S0370-2693(99)01355-6} {\bibfield  {journal}
  {\bibinfo  {journal} {Phys. Lett. B}\ }\textbf {\bibinfo {volume} {471}},\
  \bibinfo {pages} {140} (\bibinfo {year} {1999})},\ \Eprint
  {https://arxiv.org/abs/hep-th/9910095} {arXiv:hep-th/9910095} \BibitemShut
  {NoStop}%
\bibitem [{\citenamefont {Nam}(2000)}]{Nam:1999tz}%
  \BibitemOpen
  \bibfield  {author} {\bibinfo {author} {\bibfnamefont {S.}~\bibnamefont
  {Nam}},\ }\bibfield  {title} {\bibinfo {title} {{Modeling a network of brane
  worlds}},\ }\href {https://doi.org/10.1088/1126-6708/2000/03/005} {\bibfield
  {journal} {\bibinfo  {journal} {JHEP}\ }\textbf {\bibinfo {volume} {03}},\
  \bibinfo {pages} {005}},\ \Eprint {https://arxiv.org/abs/hep-th/9911104}
  {arXiv:hep-th/9911104} \BibitemShut {NoStop}%
\bibitem [{\citenamefont {Bazeia}\ and\ \citenamefont
  {Brito}(2000{\natexlab{b}})}]{Bazeia:1999xi}%
  \BibitemOpen
  \bibfield  {author} {\bibinfo {author} {\bibfnamefont {D.}~\bibnamefont
  {Bazeia}}\ and\ \bibinfo {author} {\bibfnamefont {F.~A.}\ \bibnamefont
  {Brito}},\ }\bibfield  {title} {\bibinfo {title} {{Bags, junctions, and
  networks of BPS and nonBPS defects}},\ }\href
  {https://doi.org/10.1103/PhysRevD.61.105019} {\bibfield  {journal} {\bibinfo
  {journal} {Phys. Rev. D}\ }\textbf {\bibinfo {volume} {61}},\ \bibinfo
  {pages} {105019} (\bibinfo {year} {2000}{\natexlab{b}})},\ \Eprint
  {https://arxiv.org/abs/hep-th/9912015} {arXiv:hep-th/9912015} \BibitemShut
  {NoStop}%
\bibitem [{\citenamefont {Eto}\ \emph {et~al.}(2005{\natexlab{a}})\citenamefont
  {Eto}, \citenamefont {Isozumi}, \citenamefont {Nitta}, \citenamefont
  {Ohashi},\ and\ \citenamefont {Sakai}}]{Eto:2005cp}%
  \BibitemOpen
  \bibfield  {author} {\bibinfo {author} {\bibfnamefont {M.}~\bibnamefont
  {Eto}}, \bibinfo {author} {\bibfnamefont {Y.}~\bibnamefont {Isozumi}},
  \bibinfo {author} {\bibfnamefont {M.}~\bibnamefont {Nitta}}, \bibinfo
  {author} {\bibfnamefont {K.}~\bibnamefont {Ohashi}},\ and\ \bibinfo {author}
  {\bibfnamefont {N.}~\bibnamefont {Sakai}},\ }\bibfield  {title} {\bibinfo
  {title} {{Webs of walls}},\ }\href
  {https://doi.org/10.1103/PhysRevD.72.085004} {\bibfield  {journal} {\bibinfo
  {journal} {Phys. Rev. D}\ }\textbf {\bibinfo {volume} {72}},\ \bibinfo
  {pages} {085004} (\bibinfo {year} {2005}{\natexlab{a}})},\ \Eprint
  {https://arxiv.org/abs/hep-th/0506135} {arXiv:hep-th/0506135} \BibitemShut
  {NoStop}%
\bibitem [{\citenamefont {Eto}\ \emph {et~al.}(2006{\natexlab{a}})\citenamefont
  {Eto}, \citenamefont {Isozumi}, \citenamefont {Nitta}, \citenamefont
  {Ohashi},\ and\ \citenamefont {Sakai}}]{Eto:2005fm}%
  \BibitemOpen
  \bibfield  {author} {\bibinfo {author} {\bibfnamefont {M.}~\bibnamefont
  {Eto}}, \bibinfo {author} {\bibfnamefont {Y.}~\bibnamefont {Isozumi}},
  \bibinfo {author} {\bibfnamefont {M.}~\bibnamefont {Nitta}}, \bibinfo
  {author} {\bibfnamefont {K.}~\bibnamefont {Ohashi}},\ and\ \bibinfo {author}
  {\bibfnamefont {N.}~\bibnamefont {Sakai}},\ }\bibfield  {title} {\bibinfo
  {title} {{Non-Abelian webs of walls}},\ }\href
  {https://doi.org/10.1016/j.physletb.2005.10.017} {\bibfield  {journal}
  {\bibinfo  {journal} {Phys. Lett. B}\ }\textbf {\bibinfo {volume} {632}},\
  \bibinfo {pages} {384} (\bibinfo {year} {2006}{\natexlab{a}})},\ \Eprint
  {https://arxiv.org/abs/hep-th/0508241} {arXiv:hep-th/0508241} \BibitemShut
  {NoStop}%
\bibitem [{\citenamefont {Eto}\ \emph {et~al.}(2006{\natexlab{b}})\citenamefont
  {Eto}, \citenamefont {Isozumi}, \citenamefont {Nitta}, \citenamefont
  {Ohashi},\ and\ \citenamefont {Sakai}}]{Eto:2006pg}%
  \BibitemOpen
  \bibfield  {author} {\bibinfo {author} {\bibfnamefont {M.}~\bibnamefont
  {Eto}}, \bibinfo {author} {\bibfnamefont {Y.}~\bibnamefont {Isozumi}},
  \bibinfo {author} {\bibfnamefont {M.}~\bibnamefont {Nitta}}, \bibinfo
  {author} {\bibfnamefont {K.}~\bibnamefont {Ohashi}},\ and\ \bibinfo {author}
  {\bibfnamefont {N.}~\bibnamefont {Sakai}},\ }\bibfield  {title} {\bibinfo
  {title} {{Solitons in the Higgs phase: The Moduli matrix approach}},\ }\href
  {https://doi.org/10.1088/0305-4470/39/26/R01} {\bibfield  {journal} {\bibinfo
   {journal} {J. Phys. A}\ }\textbf {\bibinfo {volume} {39}},\ \bibinfo {pages}
  {R315} (\bibinfo {year} {2006}{\natexlab{b}})},\ \Eprint
  {https://arxiv.org/abs/hep-th/0602170} {arXiv:hep-th/0602170} \BibitemShut
  {NoStop}%
\bibitem [{\citenamefont {Eto}\ \emph {et~al.}(2020{\natexlab{a}})\citenamefont
  {Eto}, \citenamefont {Kawaguchi}, \citenamefont {Nitta},\ and\ \citenamefont
  {Sasaki}}]{Eto:2020vjm}%
  \BibitemOpen
  \bibfield  {author} {\bibinfo {author} {\bibfnamefont {M.}~\bibnamefont
  {Eto}}, \bibinfo {author} {\bibfnamefont {M.}~\bibnamefont {Kawaguchi}},
  \bibinfo {author} {\bibfnamefont {M.}~\bibnamefont {Nitta}},\ and\ \bibinfo
  {author} {\bibfnamefont {R.}~\bibnamefont {Sasaki}},\ }\bibfield  {title}
  {\bibinfo {title} {{Exact solutions of domain wall junctions in arbitrary
  dimensions}},\ }\href {https://doi.org/10.1103/PhysRevD.102.065006}
  {\bibfield  {journal} {\bibinfo  {journal} {Phys. Rev. D}\ }\textbf {\bibinfo
  {volume} {102}},\ \bibinfo {pages} {065006} (\bibinfo {year}
  {2020}{\natexlab{a}})},\ \Eprint {https://arxiv.org/abs/2001.07552}
  {arXiv:2001.07552 [hep-th]} \BibitemShut {NoStop}%
\bibitem [{\citenamefont {Eto}\ \emph {et~al.}(2020{\natexlab{b}})\citenamefont
  {Eto}, \citenamefont {Kawaguchi}, \citenamefont {Nitta},\ and\ \citenamefont
  {Sasaki}}]{Eto:2020cys}%
  \BibitemOpen
  \bibfield  {author} {\bibinfo {author} {\bibfnamefont {M.}~\bibnamefont
  {Eto}}, \bibinfo {author} {\bibfnamefont {M.}~\bibnamefont {Kawaguchi}},
  \bibinfo {author} {\bibfnamefont {M.}~\bibnamefont {Nitta}},\ and\ \bibinfo
  {author} {\bibfnamefont {R.}~\bibnamefont {Sasaki}},\ }\bibfield  {title}
  {\bibinfo {title} {{Exhausting all exact solutions of BPS domain wall
  networks in arbitrary dimensions}},\ }\href
  {https://doi.org/10.1103/PhysRevD.101.105020} {\bibfield  {journal} {\bibinfo
   {journal} {Phys. Rev. D}\ }\textbf {\bibinfo {volume} {101}},\ \bibinfo
  {pages} {105020} (\bibinfo {year} {2020}{\natexlab{b}})},\ \Eprint
  {https://arxiv.org/abs/2003.13520} {arXiv:2003.13520 [hep-th]} \BibitemShut
  {NoStop}%
\bibitem [{\citenamefont {Shifman}\ and\ \citenamefont
  {Yung}(2023)}]{Shifman:2009zz}%
  \BibitemOpen
  \bibfield  {author} {\bibinfo {author} {\bibfnamefont {M.}~\bibnamefont
  {Shifman}}\ and\ \bibinfo {author} {\bibfnamefont {A.}~\bibnamefont {Yung}},\
  }\href {https://doi.org/10.1017/9781009402200} {\emph {\bibinfo {title}
  {{Supersymmetric Solitons}}}},\ Cambridge Monographs on Mathematical Physics\
  (\bibinfo  {publisher} {Cambridge University Press},\ \bibinfo {year}
  {2023})\BibitemShut {NoStop}%
\bibitem [{\citenamefont {Kim}\ \emph {et~al.}(2018)\citenamefont {Kim},
  \citenamefont {Jeong}, \citenamefont {Chu}, \citenamefont {Lee},
  \citenamefont {Kim}, \citenamefont {Xue}, \citenamefont {Koo}, \citenamefont
  {Chen}, \citenamefont {Choi}, \citenamefont {Ramesh},\ and\ \citenamefont
  {Yang}}]{Kim2018c}%
  \BibitemOpen
  \bibfield  {author} {\bibinfo {author} {\bibfnamefont {K.-E.}\ \bibnamefont
  {Kim}}, \bibinfo {author} {\bibfnamefont {S.}~\bibnamefont {Jeong}}, \bibinfo
  {author} {\bibfnamefont {K.}~\bibnamefont {Chu}}, \bibinfo {author}
  {\bibfnamefont {J.~H.}\ \bibnamefont {Lee}}, \bibinfo {author} {\bibfnamefont
  {G.-Y.}\ \bibnamefont {Kim}}, \bibinfo {author} {\bibfnamefont
  {F.}~\bibnamefont {Xue}}, \bibinfo {author} {\bibfnamefont {T.~Y.}\
  \bibnamefont {Koo}}, \bibinfo {author} {\bibfnamefont {L.-Q.}\ \bibnamefont
  {Chen}}, \bibinfo {author} {\bibfnamefont {S.-Y.}\ \bibnamefont {Choi}},
  \bibinfo {author} {\bibfnamefont {R.}~\bibnamefont {Ramesh}},\ and\ \bibinfo
  {author} {\bibfnamefont {C.-H.}\ \bibnamefont {Yang}},\ }\bibfield  {title}
  {\bibinfo {title} {Configurable topological textures in strain graded
  ferroelectric nanoplates},\ }\href
  {https://doi.org/10.1038/s41467-017-02813-5} {\bibfield  {journal} {\bibinfo
  {journal} {Nat. Commun.}\ }\textbf {\bibinfo {volume} {9}},\ \bibinfo {pages}
  {403} (\bibinfo {year} {2018})}\BibitemShut {NoStop}%
\bibitem [{\citenamefont {Wang}\ \emph {et~al.}(2022)\citenamefont {Wang},
  \citenamefont {Ma}, \citenamefont {Huang}, \citenamefont {Ma}, \citenamefont
  {Jafri}, \citenamefont {Fan}, \citenamefont {Yang}, \citenamefont {Wang},
  \citenamefont {Chen}, \citenamefont {Liu}, \citenamefont {Zhang},
  \citenamefont {Lin}, \citenamefont {Chen}, \citenamefont {Yi},\ and\
  \citenamefont {Nan}}]{Wang2022i}%
  \BibitemOpen
  \bibfield  {author} {\bibinfo {author} {\bibfnamefont {J.}~\bibnamefont
  {Wang}}, \bibinfo {author} {\bibfnamefont {J.}~\bibnamefont {Ma}}, \bibinfo
  {author} {\bibfnamefont {H.}~\bibnamefont {Huang}}, \bibinfo {author}
  {\bibfnamefont {J.}~\bibnamefont {Ma}}, \bibinfo {author} {\bibfnamefont
  {H.~M.}\ \bibnamefont {Jafri}}, \bibinfo {author} {\bibfnamefont
  {Y.}~\bibnamefont {Fan}}, \bibinfo {author} {\bibfnamefont {H.}~\bibnamefont
  {Yang}}, \bibinfo {author} {\bibfnamefont {Y.}~\bibnamefont {Wang}}, \bibinfo
  {author} {\bibfnamefont {M.}~\bibnamefont {Chen}}, \bibinfo {author}
  {\bibfnamefont {D.}~\bibnamefont {Liu}}, \bibinfo {author} {\bibfnamefont
  {J.}~\bibnamefont {Zhang}}, \bibinfo {author} {\bibfnamefont {Y.-H.}\
  \bibnamefont {Lin}}, \bibinfo {author} {\bibfnamefont {L.-Q.}\ \bibnamefont
  {Chen}}, \bibinfo {author} {\bibfnamefont {D.}~\bibnamefont {Yi}},\ and\
  \bibinfo {author} {\bibfnamefont {C.-W.}\ \bibnamefont {Nan}},\ }\bibfield
  {title} {\bibinfo {title} {Ferroelectric domain-wall logic units},\ }\href
  {https://doi.org/10.1038/s41467-022-30983-4} {\bibfield  {journal} {\bibinfo
  {journal} {Nat. Commun.}\ }\textbf {\bibinfo {volume} {13}},\ \bibinfo
  {pages} {3255} (\bibinfo {year} {2022})}\BibitemShut {NoStop}%
\bibitem [{\citenamefont {Moriya}\ \emph {et~al.}(2008)\citenamefont {Moriya},
  \citenamefont {Thomas}, \citenamefont {Hayashi}, \citenamefont {Bazaliy},
  \citenamefont {Rettner},\ and\ \citenamefont {Parkin}}]{moriya_natphys_2008}%
  \BibitemOpen
  \bibfield  {author} {\bibinfo {author} {\bibfnamefont {R.}~\bibnamefont
  {Moriya}}, \bibinfo {author} {\bibfnamefont {L.}~\bibnamefont {Thomas}},
  \bibinfo {author} {\bibfnamefont {M.}~\bibnamefont {Hayashi}}, \bibinfo
  {author} {\bibfnamefont {Y.~B.}\ \bibnamefont {Bazaliy}}, \bibinfo {author}
  {\bibfnamefont {C.}~\bibnamefont {Rettner}},\ and\ \bibinfo {author}
  {\bibfnamefont {S.~S.~P.}\ \bibnamefont {Parkin}},\ }\bibfield  {title}
  {\bibinfo {title} {Probing vortex-core dynamics using current-induced
  resonant excitation of a trapped domain wall},\ }\href
  {https://doi.org/10.1038/nphys936} {\bibfield  {journal} {\bibinfo  {journal}
  {Nature Physics}\ }\textbf {\bibinfo {volume} {4}},\ \bibinfo {pages} {368}
  (\bibinfo {year} {2008})}\BibitemShut {NoStop}%
\bibitem [{\citenamefont {Heyne}\ \emph {et~al.}(2010)\citenamefont {Heyne},
  \citenamefont {Rhensius}, \citenamefont {Ilgaz}, \citenamefont {Bisig},
  \citenamefont {R\"udiger}, \citenamefont {Kl\"aui}, \citenamefont {Joly},
  \citenamefont {Nolting}, \citenamefont {Heyderman}, \citenamefont {Thiele},\
  and\ \citenamefont {Kronast}}]{heyne_prl_2010}%
  \BibitemOpen
  \bibfield  {author} {\bibinfo {author} {\bibfnamefont {L.}~\bibnamefont
  {Heyne}}, \bibinfo {author} {\bibfnamefont {J.}~\bibnamefont {Rhensius}},
  \bibinfo {author} {\bibfnamefont {D.}~\bibnamefont {Ilgaz}}, \bibinfo
  {author} {\bibfnamefont {A.}~\bibnamefont {Bisig}}, \bibinfo {author}
  {\bibfnamefont {U.}~\bibnamefont {R\"udiger}}, \bibinfo {author}
  {\bibfnamefont {M.}~\bibnamefont {Kl\"aui}}, \bibinfo {author} {\bibfnamefont
  {L.}~\bibnamefont {Joly}}, \bibinfo {author} {\bibfnamefont {F.}~\bibnamefont
  {Nolting}}, \bibinfo {author} {\bibfnamefont {L.~J.}\ \bibnamefont
  {Heyderman}}, \bibinfo {author} {\bibfnamefont {J.~U.}\ \bibnamefont
  {Thiele}},\ and\ \bibinfo {author} {\bibfnamefont {F.}~\bibnamefont
  {Kronast}},\ }\bibfield  {title} {\bibinfo {title} {Direct determination of
  large spin-torque nonadiabaticity in vortex core dynamics},\ }\href
  {https://doi.org/10.1103/PhysRevLett.105.187203} {\bibfield  {journal}
  {\bibinfo  {journal} {Phys. Rev. Lett.}\ }\textbf {\bibinfo {volume} {105}},\
  \bibinfo {pages} {187203} (\bibinfo {year} {2010})}\BibitemShut {NoStop}%
\bibitem [{\citenamefont {Murakami}\ \emph {et~al.}(2016)\citenamefont
  {Murakami}, \citenamefont {Suzuki}, \citenamefont {Nii}, \citenamefont
  {Murai}, \citenamefont {Arima}, \citenamefont {Kainuma},\ and\ \citenamefont
  {Shindo}}]{murakami_micro_2016}%
  \BibitemOpen
  \bibfield  {author} {\bibinfo {author} {\bibfnamefont {Y.}~\bibnamefont
  {Murakami}}, \bibinfo {author} {\bibfnamefont {T.}~\bibnamefont {Suzuki}},
  \bibinfo {author} {\bibfnamefont {Y.}~\bibnamefont {Nii}}, \bibinfo {author}
  {\bibfnamefont {S.}~\bibnamefont {Murai}}, \bibinfo {author} {\bibfnamefont
  {T.}~\bibnamefont {Arima}}, \bibinfo {author} {\bibfnamefont
  {R.}~\bibnamefont {Kainuma}},\ and\ \bibinfo {author} {\bibfnamefont
  {D.}~\bibnamefont {Shindo}},\ }\bibfield  {title} {\bibinfo {title}
  {Application of strain to orbital-spin-coupled system mnv 2 o 4 at cryogenic
  temperatures within a transmission electron microscope},\ }\href
  {https://doi.org/10.1093/jmicro/dfv377} {\bibfield  {journal} {\bibinfo
  {journal} {Microscopy}\ }\textbf {\bibinfo {volume} {65}},\ \bibinfo {pages}
  {223} (\bibinfo {year} {2016})}\BibitemShut {NoStop}%
\bibitem [{\citenamefont {Skyrme}(1961)}]{Skyrme1961}%
  \BibitemOpen
  \bibfield  {author} {\bibinfo {author} {\bibfnamefont {T.~H.~R.}\
  \bibnamefont {Skyrme}},\ }\bibfield  {title} {\bibinfo {title} {A non-linear
  field theory},\ }\href {https://doi.org/10.1098/rspa.1961.0018} {\bibfield
  {journal} {\bibinfo  {journal} {Proc. R. Soc. Lond. A}\ }\textbf {\bibinfo
  {volume} {260}},\ \bibinfo {pages} {127} (\bibinfo {year}
  {1961})}\BibitemShut {NoStop}%
\bibitem [{\citenamefont {Skyrme}(1962)}]{Skyrme1962}%
  \BibitemOpen
  \bibfield  {author} {\bibinfo {author} {\bibfnamefont {T.~H.~R.}\
  \bibnamefont {Skyrme}},\ }\bibfield  {title} {\bibinfo {title} {A unified
  field theory of mesons and baryons},\ }\href
  {https://doi.org/10.1016/0029-5582(62)90775-7} {\bibfield  {journal}
  {\bibinfo  {journal} {Nucl. Phys.}\ }\textbf {\bibinfo {volume} {31}},\
  \bibinfo {pages} {556} (\bibinfo {year} {1962})}\BibitemShut {NoStop}%
\bibitem [{\citenamefont {Zhao}\ \emph {et~al.}(2023)\citenamefont {Zhao},
  \citenamefont {Malomed},\ and\ \citenamefont {Smalyukh}}]{zhao_natcom_2023}%
  \BibitemOpen
  \bibfield  {author} {\bibinfo {author} {\bibfnamefont {H.}~\bibnamefont
  {Zhao}}, \bibinfo {author} {\bibfnamefont {B.~A.}\ \bibnamefont {Malomed}},\
  and\ \bibinfo {author} {\bibfnamefont {I.~I.}\ \bibnamefont {Smalyukh}},\
  }\bibfield  {title} {\bibinfo {title} {Topological solitonic
  macromolecules},\ }\href {https://doi.org/10.1038/s41467-023-40335-5}
  {\bibfield  {journal} {\bibinfo  {journal} {Nat. Commun.}\ }\textbf {\bibinfo
  {volume} {14}},\ \bibinfo {pages} {4581} (\bibinfo {year}
  {2023})}\BibitemShut {NoStop}%
\bibitem [{\citenamefont {Wu}\ and\ \citenamefont
  {Smalyukh}(2022)}]{wu_lcreview_2022}%
  \BibitemOpen
  \bibfield  {author} {\bibinfo {author} {\bibfnamefont {J.-S.}\ \bibnamefont
  {Wu}}\ and\ \bibinfo {author} {\bibfnamefont {I.~I.}\ \bibnamefont
  {Smalyukh}},\ }\bibfield  {title} {\bibinfo {title} {Hopfions, heliknotons,
  skyrmions, torons and both abelian and nonabelian vortices in chiral liquid
  crystals},\ }\href {https://doi.org/10.1080/21680396.2022.2040058} {\bibfield
   {journal} {\bibinfo  {journal} {Liq. Cryst. Rev.}\ }\textbf {\bibinfo
  {volume} {10}},\ \bibinfo {pages} {34} (\bibinfo {year} {2022})}\BibitemShut
  {NoStop}%
\bibitem [{\citenamefont {Leslie}\ \emph {et~al.}(2009)\citenamefont {Leslie},
  \citenamefont {Hansen}, \citenamefont {Wright}, \citenamefont {Deutsch},\
  and\ \citenamefont {Bigelow}}]{leslie_prl_2009}%
  \BibitemOpen
  \bibfield  {author} {\bibinfo {author} {\bibfnamefont {L.~S.}\ \bibnamefont
  {Leslie}}, \bibinfo {author} {\bibfnamefont {A.}~\bibnamefont {Hansen}},
  \bibinfo {author} {\bibfnamefont {K.~C.}\ \bibnamefont {Wright}}, \bibinfo
  {author} {\bibfnamefont {B.~M.}\ \bibnamefont {Deutsch}},\ and\ \bibinfo
  {author} {\bibfnamefont {N.~P.}\ \bibnamefont {Bigelow}},\ }\bibfield
  {title} {\bibinfo {title} {Creation and detection of skyrmions in a
  bose-einstein condensate},\ }\href
  {https://doi.org/10.1103/PhysRevLett.103.250401} {\bibfield  {journal}
  {\bibinfo  {journal} {Phys. Rev. Lett.}\ }\textbf {\bibinfo {volume} {103}},\
  \bibinfo {pages} {250401} (\bibinfo {year} {2009})}\BibitemShut {NoStop}%
\bibitem [{\citenamefont {Choi}\ \emph {et~al.}(2012)\citenamefont {Choi},
  \citenamefont {Kwon},\ and\ \citenamefont {Shin}}]{choi_prl_2012}%
  \BibitemOpen
  \bibfield  {author} {\bibinfo {author} {\bibfnamefont {J.-y.}\ \bibnamefont
  {Choi}}, \bibinfo {author} {\bibfnamefont {W.~J.}\ \bibnamefont {Kwon}},\
  and\ \bibinfo {author} {\bibfnamefont {Y.-i.}\ \bibnamefont {Shin}},\
  }\bibfield  {title} {\bibinfo {title} {Observation of topologically stable 2d
  skyrmions in an antiferromagnetic spinor bose-einstein condensate},\ }\href
  {https://doi.org/10.1103/PhysRevLett.108.035301} {\bibfield  {journal}
  {\bibinfo  {journal} {Phys. Rev. Lett.}\ }\textbf {\bibinfo {volume} {108}},\
  \bibinfo {pages} {035301} (\bibinfo {year} {2012})}\BibitemShut {NoStop}%
\bibitem [{\citenamefont {Speight}\ \emph
  {et~al.}(2023{\natexlab{a}})\citenamefont {Speight}, \citenamefont
  {Winyard},\ and\ \citenamefont {Babaev}}]{Speight2023a}%
  \BibitemOpen
  \bibfield  {author} {\bibinfo {author} {\bibfnamefont {M.}~\bibnamefont
  {Speight}}, \bibinfo {author} {\bibfnamefont {T.}~\bibnamefont {Winyard}},\
  and\ \bibinfo {author} {\bibfnamefont {E.}~\bibnamefont {Babaev}},\
  }\bibfield  {title} {\bibinfo {title} {Symmetries, length scales, magnetic
  response, and skyrmion chains in nematic superconductors},\ }\href
  {https://doi.org/10.1103/PhysRevB.107.195204} {\bibfield  {journal} {\bibinfo
   {journal} {Phys. Rev. B}\ }\textbf {\bibinfo {volume} {107}},\ \bibinfo
  {pages} {195204} (\bibinfo {year} {2023}{\natexlab{a}})}\BibitemShut
  {NoStop}%
\bibitem [{\citenamefont {Speight}\ \emph
  {et~al.}(2023{\natexlab{b}})\citenamefont {Speight}, \citenamefont
  {Winyard},\ and\ \citenamefont {Babaev}}]{Speight2023b}%
  \BibitemOpen
  \bibfield  {author} {\bibinfo {author} {\bibfnamefont {M.}~\bibnamefont
  {Speight}}, \bibinfo {author} {\bibfnamefont {T.}~\bibnamefont {Winyard}},\
  and\ \bibinfo {author} {\bibfnamefont {E.}~\bibnamefont {Babaev}},\
  }\bibfield  {title} {\bibinfo {title} {Magnetic {{Response}} of {{Nematic
  Superconductors}}: {{Skyrmion Stripes}} and {{Their Signatures}} in {{Muon
  Spin Relaxation Experiments}}},\ }\href
  {https://doi.org/10.1103/PhysRevLett.130.226002} {\bibfield  {journal}
  {\bibinfo  {journal} {Phys. Rev. Lett.}\ }\textbf {\bibinfo {volume} {130}},\
  \bibinfo {pages} {226002} (\bibinfo {year} {2023}{\natexlab{b}})}\BibitemShut
  {NoStop}%
\bibitem [{\citenamefont {Shen}\ \emph {et~al.}(2024)\citenamefont {Shen},
  \citenamefont {Zhang}, \citenamefont {Shi}, \citenamefont {Du}, \citenamefont
  {Yuan},\ and\ \citenamefont {Zayats}}]{shen_natphoto_2024}%
  \BibitemOpen
  \bibfield  {author} {\bibinfo {author} {\bibfnamefont {Y.}~\bibnamefont
  {Shen}}, \bibinfo {author} {\bibfnamefont {Q.}~\bibnamefont {Zhang}},
  \bibinfo {author} {\bibfnamefont {P.}~\bibnamefont {Shi}}, \bibinfo {author}
  {\bibfnamefont {L.}~\bibnamefont {Du}}, \bibinfo {author} {\bibfnamefont
  {X.}~\bibnamefont {Yuan}},\ and\ \bibinfo {author} {\bibfnamefont {A.~V.}\
  \bibnamefont {Zayats}},\ }\bibfield  {title} {\bibinfo {title} {Optical
  skyrmions and other topological quasiparticles of light},\ }\href
  {https://doi.org/10.1038/s41566-023-01325-7} {\bibfield  {journal} {\bibinfo
  {journal} {Nat. Photon.}\ }\textbf {\bibinfo {volume} {18}},\ \bibinfo
  {pages} {15} (\bibinfo {year} {2024})}\BibitemShut {NoStop}%
\bibitem [{\citenamefont {Bogdanov}\ and\ \citenamefont
  {Yablonskii}(1989)}]{Bogdanov:1989}%
  \BibitemOpen
  \bibfield  {author} {\bibinfo {author} {\bibfnamefont {A.}~\bibnamefont
  {Bogdanov}}\ and\ \bibinfo {author} {\bibfnamefont {D.}~\bibnamefont
  {Yablonskii}},\ }\bibfield  {title} {\bibinfo {title} {{Thermodynamically
  stable vortices in magnetically ordered crystals. The mixed state of
  magnets}},\ }\href
  {http://www.jetp.ras.ru/cgi-bin/e/index/e/68/1/p101?a=list} {\bibfield
  {journal} {\bibinfo  {journal} {Sov. Phys. JETP}\ }\textbf {\bibinfo {volume}
  {68}},\ \bibinfo {pages} {101} (\bibinfo {year} {1989})}\BibitemShut
  {NoStop}%
\bibitem [{\citenamefont {Bogdanov}(1995)}]{Bogdanov:1995}%
  \BibitemOpen
  \bibfield  {author} {\bibinfo {author} {\bibfnamefont {A.}~\bibnamefont
  {Bogdanov}},\ }\bibfield  {title} {\bibinfo {title} {{New localized solutions
  of the nonlinear field equations}},\ }\href
  {http://jetpletters.ru/ps/1214/article_18359.shtml} {\bibfield  {journal}
  {\bibinfo  {journal} {JETP Lett.}\ }\textbf {\bibinfo {volume} {62}},\
  \bibinfo {pages} {247} (\bibinfo {year} {1995})}\BibitemShut {NoStop}%
\bibitem [{\citenamefont {Speight}\ and\ \citenamefont
  {Winyard}(2020)}]{speight_prb_2020}%
  \BibitemOpen
  \bibfield  {author} {\bibinfo {author} {\bibfnamefont {M.}~\bibnamefont
  {Speight}}\ and\ \bibinfo {author} {\bibfnamefont {T.}~\bibnamefont
  {Winyard}},\ }\bibfield  {title} {\bibinfo {title} {Skyrmions and spin waves
  in frustrated ferromagnets at low applied magnetic field},\ }\href
  {https://doi.org/10.1103/PhysRevB.101.134420} {\bibfield  {journal} {\bibinfo
   {journal} {Phys. Rev. B}\ }\textbf {\bibinfo {volume} {101}},\ \bibinfo
  {pages} {134420} (\bibinfo {year} {2020})}\BibitemShut {NoStop}%
\bibitem [{\citenamefont {Göbel}\ \emph {et~al.}(2021)\citenamefont {Göbel},
  \citenamefont {Mertig},\ and\ \citenamefont
  {Tretiakov}}]{gobel_physreport_2021}%
  \BibitemOpen
  \bibfield  {author} {\bibinfo {author} {\bibfnamefont {B.}~\bibnamefont
  {Göbel}}, \bibinfo {author} {\bibfnamefont {I.}~\bibnamefont {Mertig}},\
  and\ \bibinfo {author} {\bibfnamefont {O.~A.}\ \bibnamefont {Tretiakov}},\
  }\bibfield  {title} {\bibinfo {title} {Beyond skyrmions: Review and
  perspectives of alternative magnetic quasiparticles},\ }\href
  {https://doi.org/https://doi.org/10.1016/j.physrep.2020.10.001} {\bibfield
  {journal} {\bibinfo  {journal} {Phys. Rep.}\ }\textbf {\bibinfo {volume}
  {895}},\ \bibinfo {pages} {1} (\bibinfo {year} {2021})}\BibitemShut {NoStop}%
\bibitem [{\citenamefont {Rossler}\ \emph {et~al.}(2006)\citenamefont
  {Rossler}, \citenamefont {Bogdanov},\ and\ \citenamefont
  {Pfleiderer}}]{Rossler:2006}%
  \BibitemOpen
  \bibfield  {author} {\bibinfo {author} {\bibfnamefont {U.~K.}\ \bibnamefont
  {Rossler}}, \bibinfo {author} {\bibfnamefont {A.~N.}\ \bibnamefont
  {Bogdanov}},\ and\ \bibinfo {author} {\bibfnamefont {C.}~\bibnamefont
  {Pfleiderer}},\ }\bibfield  {title} {\bibinfo {title} {Spontaneous skyrmion
  ground states in magnetic metals},\ }\href
  {https://doi.org/10.1038/nature05056} {\bibfield  {journal} {\bibinfo
  {journal} {Nature}\ }\textbf {\bibinfo {volume} {442}},\ \bibinfo {pages}
  {797} (\bibinfo {year} {2006})}\BibitemShut {NoStop}%
\bibitem [{\citenamefont {M\"{u}hlbauer}\ \emph {et~al.}(2009)\citenamefont
  {M\"{u}hlbauer}, \citenamefont {Binz}, \citenamefont {Jonietz}, \citenamefont
  {Pfleiderer}, \citenamefont {Rosch}, \citenamefont {Neubauer}, \citenamefont
  {Georgii},\ and\ \citenamefont {Boni}}]{doi:10.1126/science.1166767}%
  \BibitemOpen
  \bibfield  {author} {\bibinfo {author} {\bibfnamefont {S.}~\bibnamefont
  {M\"{u}hlbauer}}, \bibinfo {author} {\bibfnamefont {B.}~\bibnamefont {Binz}},
  \bibinfo {author} {\bibfnamefont {F.}~\bibnamefont {Jonietz}}, \bibinfo
  {author} {\bibfnamefont {C.}~\bibnamefont {Pfleiderer}}, \bibinfo {author}
  {\bibfnamefont {A.}~\bibnamefont {Rosch}}, \bibinfo {author} {\bibfnamefont
  {A.}~\bibnamefont {Neubauer}}, \bibinfo {author} {\bibfnamefont
  {R.}~\bibnamefont {Georgii}},\ and\ \bibinfo {author} {\bibfnamefont
  {P.}~\bibnamefont {Boni}},\ }\bibfield  {title} {\bibinfo {title} {Skyrmion
  lattice in a chiral magnet},\ }\href
  {https://doi.org/10.1126/science.1166767} {\bibfield  {journal} {\bibinfo
  {journal} {Science}\ }\textbf {\bibinfo {volume} {323}},\ \bibinfo {pages}
  {915} (\bibinfo {year} {2009})}\BibitemShut {NoStop}%
\bibitem [{\citenamefont {Yu}\ \emph {et~al.}(2010)\citenamefont {Yu},
  \citenamefont {Onose}, \citenamefont {Kanazawa}, \citenamefont {Park},
  \citenamefont {Han}, \citenamefont {Matsui}, \citenamefont {Nagaosa},\ and\
  \citenamefont {Tokura}}]{doi:10.1038/nature09124}%
  \BibitemOpen
  \bibfield  {author} {\bibinfo {author} {\bibfnamefont {X.~Z.}\ \bibnamefont
  {Yu}}, \bibinfo {author} {\bibfnamefont {Y.}~\bibnamefont {Onose}}, \bibinfo
  {author} {\bibfnamefont {N.}~\bibnamefont {Kanazawa}}, \bibinfo {author}
  {\bibfnamefont {J.~H.}\ \bibnamefont {Park}}, \bibinfo {author}
  {\bibfnamefont {J.~H.}\ \bibnamefont {Han}}, \bibinfo {author} {\bibfnamefont
  {Y.}~\bibnamefont {Matsui}}, \bibinfo {author} {\bibfnamefont
  {N.}~\bibnamefont {Nagaosa}},\ and\ \bibinfo {author} {\bibfnamefont
  {Y.}~\bibnamefont {Tokura}},\ }\bibfield  {title} {\bibinfo {title}
  {{Real-space observation of a two-dimensional skyrmion crystal}},\ }\href
  {https://doi.org/10.1038/nature09124} {\bibfield  {journal} {\bibinfo
  {journal} {Nature}\ }\textbf {\bibinfo {volume} {465}},\ \bibinfo {pages}
  {901} (\bibinfo {year} {2010})}\BibitemShut {NoStop}%
\bibitem [{\citenamefont {Han}\ \emph {et~al.}(2010)\citenamefont {Han},
  \citenamefont {Zang}, \citenamefont {Yang}, \citenamefont {Park},\ and\
  \citenamefont {Nagaosa}}]{Han:2010by}%
  \BibitemOpen
  \bibfield  {author} {\bibinfo {author} {\bibfnamefont {J.~H.}\ \bibnamefont
  {Han}}, \bibinfo {author} {\bibfnamefont {J.}~\bibnamefont {Zang}}, \bibinfo
  {author} {\bibfnamefont {Z.}~\bibnamefont {Yang}}, \bibinfo {author}
  {\bibfnamefont {J.-H.}\ \bibnamefont {Park}},\ and\ \bibinfo {author}
  {\bibfnamefont {N.}~\bibnamefont {Nagaosa}},\ }\bibfield  {title} {\bibinfo
  {title} {{Skyrmion Lattice in Two-Dimensional Chiral Magnet}},\ }\href
  {https://doi.org/10.1103/PhysRevB.82.094429} {\bibfield  {journal} {\bibinfo
  {journal} {Phys. Rev. B}\ }\textbf {\bibinfo {volume} {82}},\ \bibinfo
  {pages} {094429} (\bibinfo {year} {2010})}\BibitemShut {NoStop}%
\bibitem [{\citenamefont {Heinze}\ \emph {et~al.}(2011)\citenamefont {Heinze},
  \citenamefont {von Bergmann}, \citenamefont {Menzel}, \citenamefont {Brede},
  \citenamefont {Kubetzka}, \citenamefont {Wiesendanger}, \citenamefont
  {Bihlmayer},\ and\ \citenamefont {Blugel}}]{doi:10.1038/nphys2045}%
  \BibitemOpen
  \bibfield  {author} {\bibinfo {author} {\bibfnamefont {S.}~\bibnamefont
  {Heinze}}, \bibinfo {author} {\bibfnamefont {K.}~\bibnamefont {von
  Bergmann}}, \bibinfo {author} {\bibfnamefont {M.}~\bibnamefont {Menzel}},
  \bibinfo {author} {\bibfnamefont {J.}~\bibnamefont {Brede}}, \bibinfo
  {author} {\bibfnamefont {A.}~\bibnamefont {Kubetzka}}, \bibinfo {author}
  {\bibfnamefont {R.}~\bibnamefont {Wiesendanger}}, \bibinfo {author}
  {\bibfnamefont {G.}~\bibnamefont {Bihlmayer}},\ and\ \bibinfo {author}
  {\bibfnamefont {S.}~\bibnamefont {Blugel}},\ }\bibfield  {title} {\bibinfo
  {title} {{Spontaneous atomic-scale magnetic skyrmion lattice in two
  dimensions}},\ }\href {https://doi.org/10.1038/nphys2045} {\bibfield
  {journal} {\bibinfo  {journal} {Nat. Phys.}\ }\textbf {\bibinfo {volume}
  {7}},\ \bibinfo {pages} {713} (\bibinfo {year} {2011})}\BibitemShut {NoStop}%
\bibitem [{\citenamefont {Nagaosa}\ and\ \citenamefont
  {Tokura}(2013)}]{Nagaosa2013}%
  \BibitemOpen
  \bibfield  {author} {\bibinfo {author} {\bibfnamefont {N.}~\bibnamefont
  {Nagaosa}}\ and\ \bibinfo {author} {\bibfnamefont {Y.}~\bibnamefont
  {Tokura}},\ }\bibfield  {title} {\bibinfo {title} {{Topological properties
  and dynamics of magnetic skyrmions}},\ }\href
  {https://doi.org/10.1038/nnano.2013.243} {\bibfield  {journal} {\bibinfo
  {journal} {Nat. Nanotechnol.}\ }\textbf {\bibinfo {volume} {8}},\ \bibinfo
  {pages} {899} (\bibinfo {year} {2013})}\BibitemShut {NoStop}%
\bibitem [{\citenamefont {Eto}\ \emph {et~al.}(2005{\natexlab{b}})\citenamefont
  {Eto}, \citenamefont {Nitta}, \citenamefont {Ohashi},\ and\ \citenamefont
  {Tong}}]{Eto2005c}%
  \BibitemOpen
  \bibfield  {author} {\bibinfo {author} {\bibfnamefont {M.}~\bibnamefont
  {Eto}}, \bibinfo {author} {\bibfnamefont {M.}~\bibnamefont {Nitta}}, \bibinfo
  {author} {\bibfnamefont {K.}~\bibnamefont {Ohashi}},\ and\ \bibinfo {author}
  {\bibfnamefont {D.}~\bibnamefont {Tong}},\ }\bibfield  {title} {\bibinfo
  {title} {Skyrmions from {{Instantons}} inside {{Domain Walls}}},\ }\href
  {https://doi.org/10.1103/PhysRevLett.95.252003} {\bibfield  {journal}
  {\bibinfo  {journal} {Phys. Rev. Lett.}\ }\textbf {\bibinfo {volume} {95}},\
  \bibinfo {pages} {252003} (\bibinfo {year} {2005}{\natexlab{b}})}\BibitemShut
  {NoStop}%
\bibitem [{\citenamefont {Nitta}(2012)}]{Nitta:2012xq}%
  \BibitemOpen
  \bibfield  {author} {\bibinfo {author} {\bibfnamefont {M.}~\bibnamefont
  {Nitta}},\ }\bibfield  {title} {\bibinfo {title} {{Josephson vortices and the
  Atiyah-Manton construction}},\ }\href
  {https://doi.org/10.1103/PhysRevD.86.125004} {\bibfield  {journal} {\bibinfo
  {journal} {Phys. Rev. D}\ }\textbf {\bibinfo {volume} {86}},\ \bibinfo
  {pages} {125004} (\bibinfo {year} {2012})},\ \Eprint
  {https://arxiv.org/abs/1207.6958} {arXiv:1207.6958 [hep-th]} \BibitemShut
  {NoStop}%
\bibitem [{\citenamefont {Kobayashi}\ and\ \citenamefont
  {Nitta}(2013)}]{Kobayashi:2013ju}%
  \BibitemOpen
  \bibfield  {author} {\bibinfo {author} {\bibfnamefont {M.}~\bibnamefont
  {Kobayashi}}\ and\ \bibinfo {author} {\bibfnamefont {M.}~\bibnamefont
  {Nitta}},\ }\bibfield  {title} {\bibinfo {title} {{Sine-Gordon kinks on a
  domain wall ring}},\ }\href {https://doi.org/10.1103/PhysRevD.87.085003}
  {\bibfield  {journal} {\bibinfo  {journal} {Phys. Rev. D}\ }\textbf {\bibinfo
  {volume} {87}},\ \bibinfo {pages} {085003} (\bibinfo {year} {2013})},\
  \Eprint {https://arxiv.org/abs/1302.0989} {arXiv:1302.0989 [hep-th]}
  \BibitemShut {NoStop}%
\bibitem [{\citenamefont {Jennings}\ and\ \citenamefont
  {Sutcliffe}(2013)}]{Jennings2013}%
  \BibitemOpen
  \bibfield  {author} {\bibinfo {author} {\bibfnamefont {P.}~\bibnamefont
  {Jennings}}\ and\ \bibinfo {author} {\bibfnamefont {P.}~\bibnamefont
  {Sutcliffe}},\ }\bibfield  {title} {\bibinfo {title} {The dynamics of domain
  wall {{Skyrmions}}},\ }\href {https://doi.org/10.1088/1751-8113/46/46/465401}
  {\bibfield  {journal} {\bibinfo  {journal} {J. Phys. A: Math. Theor.}\
  }\textbf {\bibinfo {volume} {46}},\ \bibinfo {pages} {465401} (\bibinfo
  {year} {2013})}\BibitemShut {NoStop}%
\bibitem [{\citenamefont {Nitta}(2013{\natexlab{a}})}]{Nitta:2012wi}%
  \BibitemOpen
  \bibfield  {author} {\bibinfo {author} {\bibfnamefont {M.}~\bibnamefont
  {Nitta}},\ }\bibfield  {title} {\bibinfo {title} {{Correspondence between
  Skyrmions in 2+1 and 3+1 Dimensions}},\ }\href
  {https://doi.org/10.1103/PhysRevD.87.025013} {\bibfield  {journal} {\bibinfo
  {journal} {Phys. Rev. D}\ }\textbf {\bibinfo {volume} {87}},\ \bibinfo
  {pages} {025013} (\bibinfo {year} {2013}{\natexlab{a}})},\ \Eprint
  {https://arxiv.org/abs/1210.2233} {arXiv:1210.2233 [hep-th]} \BibitemShut
  {NoStop}%
\bibitem [{\citenamefont {Nitta}(2013{\natexlab{b}})}]{Nitta2013}%
  \BibitemOpen
  \bibfield  {author} {\bibinfo {author} {\bibfnamefont {M.}~\bibnamefont
  {Nitta}},\ }\bibfield  {title} {\bibinfo {title} {Matryoshka {{Skyrmions}}},\
  }\href {https://doi.org/10.1016/j.nuclphysb.2013.03.003} {\bibfield
  {journal} {\bibinfo  {journal} {Nucl. Phys. B}\ }\textbf {\bibinfo {volume}
  {872}},\ \bibinfo {pages} {62} (\bibinfo {year}
  {2013}{\natexlab{b}})}\BibitemShut {NoStop}%
\bibitem [{\citenamefont {Gudnason}\ and\ \citenamefont
  {Nitta}(2014{\natexlab{a}})}]{Gudnason2014b}%
  \BibitemOpen
  \bibfield  {author} {\bibinfo {author} {\bibfnamefont {S.~B.}\ \bibnamefont
  {Gudnason}}\ and\ \bibinfo {author} {\bibfnamefont {M.}~\bibnamefont
  {Nitta}},\ }\bibfield  {title} {\bibinfo {title} {Domain wall
  {{Skyrmions}}},\ }\href {https://doi.org/10.1103/PhysRevD.89.085022}
  {\bibfield  {journal} {\bibinfo  {journal} {Phys. Rev. D}\ }\textbf {\bibinfo
  {volume} {89}},\ \bibinfo {pages} {085022} (\bibinfo {year}
  {2014}{\natexlab{a}})},\ \Eprint {https://arxiv.org/abs/1403.1245}
  {arXiv:1403.1245 [hep-th]} \BibitemShut {NoStop}%
\bibitem [{\citenamefont {Gudnason}\ and\ \citenamefont
  {Nitta}(2014{\natexlab{b}})}]{Gudnason:2014hsa}%
  \BibitemOpen
  \bibfield  {author} {\bibinfo {author} {\bibfnamefont {S.~B.}\ \bibnamefont
  {Gudnason}}\ and\ \bibinfo {author} {\bibfnamefont {M.}~\bibnamefont
  {Nitta}},\ }\bibfield  {title} {\bibinfo {title} {{Incarnations of
  Skyrmions}},\ }\href {https://doi.org/10.1103/PhysRevD.90.085007} {\bibfield
  {journal} {\bibinfo  {journal} {Phys. Rev. D}\ }\textbf {\bibinfo {volume}
  {90}},\ \bibinfo {pages} {085007} (\bibinfo {year} {2014}{\natexlab{b}})},\
  \Eprint {https://arxiv.org/abs/1407.7210} {arXiv:1407.7210 [hep-th]}
  \BibitemShut {NoStop}%
\bibitem [{\citenamefont {Eto}\ \emph {et~al.}(2023{\natexlab{a}})\citenamefont
  {Eto}, \citenamefont {Nishimura},\ and\ \citenamefont {Nitta}}]{Eto:2023lyo}%
  \BibitemOpen
  \bibfield  {author} {\bibinfo {author} {\bibfnamefont {M.}~\bibnamefont
  {Eto}}, \bibinfo {author} {\bibfnamefont {K.}~\bibnamefont {Nishimura}},\
  and\ \bibinfo {author} {\bibfnamefont {M.}~\bibnamefont {Nitta}},\ }\bibfield
   {title} {\bibinfo {title} {{How baryons appear in low-energy QCD:
  Domain-wall Skyrmion phase in strong magnetic fields}},\ }\href@noop {} {\
  (\bibinfo {year} {2023}{\natexlab{a}})},\ \Eprint
  {https://arxiv.org/abs/2304.02940} {arXiv:2304.02940 [hep-ph]} \BibitemShut
  {NoStop}%
\bibitem [{\citenamefont {Eto}\ \emph {et~al.}(2023{\natexlab{b}})\citenamefont
  {Eto}, \citenamefont {Nishimura},\ and\ \citenamefont {Nitta}}]{Eto:2023wul}%
  \BibitemOpen
  \bibfield  {author} {\bibinfo {author} {\bibfnamefont {M.}~\bibnamefont
  {Eto}}, \bibinfo {author} {\bibfnamefont {K.}~\bibnamefont {Nishimura}},\
  and\ \bibinfo {author} {\bibfnamefont {M.}~\bibnamefont {Nitta}},\ }\bibfield
   {title} {\bibinfo {title} {{Phase diagram of QCD matter with magnetic field:
  domain-wall Skyrmion chain in chiral soliton lattice}},\ }\href
  {https://doi.org/10.1007/JHEP12(2023)032} {\bibfield  {journal} {\bibinfo
  {journal} {JHEP}\ }\textbf {\bibinfo {volume} {12}},\ \bibinfo {pages}
  {032}},\ \Eprint {https://arxiv.org/abs/2311.01112} {arXiv:2311.01112
  [hep-ph]} \BibitemShut {NoStop}%
\bibitem [{\citenamefont {Eto}\ \emph {et~al.}(2024)\citenamefont {Eto},
  \citenamefont {Nishimura},\ and\ \citenamefont {Nitta}}]{Eto:2023tuu}%
  \BibitemOpen
  \bibfield  {author} {\bibinfo {author} {\bibfnamefont {M.}~\bibnamefont
  {Eto}}, \bibinfo {author} {\bibfnamefont {K.}~\bibnamefont {Nishimura}},\
  and\ \bibinfo {author} {\bibfnamefont {M.}~\bibnamefont {Nitta}},\ }\bibfield
   {title} {\bibinfo {title} {{Domain-wall Skyrmion phase in a rapidly rotating
  QCD matter}},\ }\href {https://doi.org/10.1007/JHEP03(2024)019} {\bibfield
  {journal} {\bibinfo  {journal} {JHEP}\ }\textbf {\bibinfo {volume} {03}},\
  \bibinfo {pages} {019}},\ \Eprint {https://arxiv.org/abs/2310.17511}
  {arXiv:2310.17511 [hep-ph]} \BibitemShut {NoStop}%
\bibitem [{\citenamefont {Amari}\ \emph
  {et~al.}(2024{\natexlab{b}})\citenamefont {Amari}, \citenamefont {Nitta},\
  and\ \citenamefont {Yokokura}}]{Amari:2024mip}%
  \BibitemOpen
  \bibfield  {author} {\bibinfo {author} {\bibfnamefont {Y.}~\bibnamefont
  {Amari}}, \bibinfo {author} {\bibfnamefont {M.}~\bibnamefont {Nitta}},\ and\
  \bibinfo {author} {\bibfnamefont {R.}~\bibnamefont {Yokokura}},\ }\bibfield
  {title} {\bibinfo {title} {{Spin Statistics and Surgeries of Topological
  Solitons in QCD Matter in Magnetic Field}},\ }\href@noop {} {\  (\bibinfo
  {year} {2024}{\natexlab{b}})},\ \Eprint {https://arxiv.org/abs/2406.14419}
  {arXiv:2406.14419 [hep-th]} \BibitemShut {NoStop}%
\bibitem [{\citenamefont {Kim}\ and\ \citenamefont
  {Tserkovnyak}(2017)}]{Kim:2017lsi}%
  \BibitemOpen
  \bibfield  {author} {\bibinfo {author} {\bibfnamefont {S.~K.}\ \bibnamefont
  {Kim}}\ and\ \bibinfo {author} {\bibfnamefont {Y.}~\bibnamefont
  {Tserkovnyak}},\ }\bibfield  {title} {\bibinfo {title} {{Magnetic Domain
  Walls as Hosts of Spin Superfluids and Generators of Skyrmions}},\ }\href
  {https://doi.org/10.1103/PhysRevLett.119.047202} {\bibfield  {journal}
  {\bibinfo  {journal} {Phys. Rev. Lett.}\ }\textbf {\bibinfo {volume} {119}},\
  \bibinfo {pages} {047202} (\bibinfo {year} {2017})},\ \Eprint
  {https://arxiv.org/abs/1701.08273} {arXiv:1701.08273 [cond-mat.mes-hall]}
  \BibitemShut {NoStop}%
\bibitem [{\citenamefont {Cheng}\ \emph {et~al.}(2019)\citenamefont {Cheng},
  \citenamefont {Li}, \citenamefont {Sapkota}, \citenamefont {Rai},
  \citenamefont {Pokhrel}, \citenamefont {Mewes}, \citenamefont {Mewes},
  \citenamefont {Xiao}, \citenamefont {De~Graef},\ and\ \citenamefont
  {Sokalski}}]{PhysRevB.99.184412}%
  \BibitemOpen
  \bibfield  {author} {\bibinfo {author} {\bibfnamefont {R.}~\bibnamefont
  {Cheng}}, \bibinfo {author} {\bibfnamefont {M.}~\bibnamefont {Li}}, \bibinfo
  {author} {\bibfnamefont {A.}~\bibnamefont {Sapkota}}, \bibinfo {author}
  {\bibfnamefont {A.}~\bibnamefont {Rai}}, \bibinfo {author} {\bibfnamefont
  {A.}~\bibnamefont {Pokhrel}}, \bibinfo {author} {\bibfnamefont
  {T.}~\bibnamefont {Mewes}}, \bibinfo {author} {\bibfnamefont
  {C.}~\bibnamefont {Mewes}}, \bibinfo {author} {\bibfnamefont
  {D.}~\bibnamefont {Xiao}}, \bibinfo {author} {\bibfnamefont {M.}~\bibnamefont
  {De~Graef}},\ and\ \bibinfo {author} {\bibfnamefont {V.}~\bibnamefont
  {Sokalski}},\ }\bibfield  {title} {\bibinfo {title} {Magnetic domain wall
  skyrmions},\ }\href {https://doi.org/10.1103/PhysRevB.99.184412} {\bibfield
  {journal} {\bibinfo  {journal} {Phys. Rev. B}\ }\textbf {\bibinfo {volume}
  {99}},\ \bibinfo {pages} {184412} (\bibinfo {year} {2019})}\BibitemShut
  {NoStop}%
\bibitem [{\citenamefont {Lepadatu}(2020)}]{PhysRevB.102.094402}%
  \BibitemOpen
  \bibfield  {author} {\bibinfo {author} {\bibfnamefont {S.}~\bibnamefont
  {Lepadatu}},\ }\bibfield  {title} {\bibinfo {title} {Emergence of transient
  domain wall skyrmions after ultrafast demagnetization},\ }\href
  {https://doi.org/10.1103/PhysRevB.102.094402} {\bibfield  {journal} {\bibinfo
   {journal} {Phys. Rev. B}\ }\textbf {\bibinfo {volume} {102}},\ \bibinfo
  {pages} {094402} (\bibinfo {year} {2020})}\BibitemShut {NoStop}%
\bibitem [{\citenamefont {Kuchkin}\ \emph {et~al.}(2020)\citenamefont
  {Kuchkin}, \citenamefont {Barton-Singer}, \citenamefont {Rybakov},
  \citenamefont {Bl\"ugel}, \citenamefont {Schroers},\ and\ \citenamefont
  {Kiselev}}]{KBRBSK}%
  \BibitemOpen
  \bibfield  {author} {\bibinfo {author} {\bibfnamefont {V.~M.}\ \bibnamefont
  {Kuchkin}}, \bibinfo {author} {\bibfnamefont {B.}~\bibnamefont
  {Barton-Singer}}, \bibinfo {author} {\bibfnamefont {F.~N.}\ \bibnamefont
  {Rybakov}}, \bibinfo {author} {\bibfnamefont {S.}~\bibnamefont {Bl\"ugel}},
  \bibinfo {author} {\bibfnamefont {B.~J.}\ \bibnamefont {Schroers}},\ and\
  \bibinfo {author} {\bibfnamefont {N.~S.}\ \bibnamefont {Kiselev}},\
  }\bibfield  {title} {\bibinfo {title} {{Magnetic skyrmions, chiral kinks and
  holomorphic functions}},\ }\href
  {https://doi.org/10.1103/PhysRevB.102.144422} {\bibfield  {journal} {\bibinfo
   {journal} {Phys. Rev. B}\ }\textbf {\bibinfo {volume} {102}},\ \bibinfo
  {pages} {144422} (\bibinfo {year} {2020})}\BibitemShut {NoStop}%
\bibitem [{\citenamefont {Ross}\ and\ \citenamefont {Nitta}(2023)}]{Ross2023}%
  \BibitemOpen
  \bibfield  {author} {\bibinfo {author} {\bibfnamefont {C.}~\bibnamefont
  {Ross}}\ and\ \bibinfo {author} {\bibfnamefont {M.}~\bibnamefont {Nitta}},\
  }\bibfield  {title} {\bibinfo {title} {Domain-wall skyrmions in chiral
  magnets},\ }\href {https://doi.org/10.1103/PhysRevB.107.024422} {\bibfield
  {journal} {\bibinfo  {journal} {Phys. Rev. B}\ }\textbf {\bibinfo {volume}
  {107}},\ \bibinfo {pages} {024422} (\bibinfo {year} {2023})}\BibitemShut
  {NoStop}%
\bibitem [{\citenamefont {Amari}\ and\ \citenamefont
  {Nitta}(2023)}]{Amari:2023gqv}%
  \BibitemOpen
  \bibfield  {author} {\bibinfo {author} {\bibfnamefont {Y.}~\bibnamefont
  {Amari}}\ and\ \bibinfo {author} {\bibfnamefont {M.}~\bibnamefont {Nitta}},\
  }\bibfield  {title} {\bibinfo {title} {{Chiral magnets from string theory}},\
  }\href {https://doi.org/10.1007/JHEP11(2023)212} {\bibfield  {journal}
  {\bibinfo  {journal} {JHEP}\ }\textbf {\bibinfo {volume} {11}},\ \bibinfo
  {pages} {212}},\ \Eprint {https://arxiv.org/abs/2307.11113} {arXiv:2307.11113
  [hep-th]} \BibitemShut {NoStop}%
\bibitem [{\citenamefont {Lee}\ \emph {et~al.}(2023)\citenamefont {Lee},
  \citenamefont {Nakata}, \citenamefont {Tchernyshyov},\ and\ \citenamefont
  {Kim}}]{Lee2023}%
  \BibitemOpen
  \bibfield  {author} {\bibinfo {author} {\bibfnamefont {S.}~\bibnamefont
  {Lee}}, \bibinfo {author} {\bibfnamefont {K.}~\bibnamefont {Nakata}},
  \bibinfo {author} {\bibfnamefont {O.}~\bibnamefont {Tchernyshyov}},\ and\
  \bibinfo {author} {\bibfnamefont {S.~K.}\ \bibnamefont {Kim}},\ }\bibfield
  {title} {\bibinfo {title} {Magnon dynamics in a skyrmion-textured domain wall
  of antiferromagnets},\ }\href {https://doi.org/10.1103/PhysRevB.107.184432}
  {\bibfield  {journal} {\bibinfo  {journal} {Phys. Rev. B}\ }\textbf {\bibinfo
  {volume} {107}},\ \bibinfo {pages} {184432} (\bibinfo {year}
  {2023})}\BibitemShut {NoStop}%
\bibitem [{\citenamefont {Amari}\ \emph
  {et~al.}(2024{\natexlab{c}})\citenamefont {Amari}, \citenamefont {Ross},\
  and\ \citenamefont {Nitta}}]{Amari2024a}%
  \BibitemOpen
  \bibfield  {author} {\bibinfo {author} {\bibfnamefont {Y.}~\bibnamefont
  {Amari}}, \bibinfo {author} {\bibfnamefont {C.}~\bibnamefont {Ross}},\ and\
  \bibinfo {author} {\bibfnamefont {M.}~\bibnamefont {Nitta}},\ }\bibfield
  {title} {\bibinfo {title} {Domain-wall skyrmion chain and domain-wall
  bimerons in chiral magnets},\ }\href
  {https://doi.org/10.1103/PhysRevB.109.104426} {\bibfield  {journal} {\bibinfo
   {journal} {Phys. Rev. B}\ }\textbf {\bibinfo {volume} {109}},\ \bibinfo
  {pages} {104426} (\bibinfo {year} {2024}{\natexlab{c}})}\BibitemShut
  {NoStop}%
\bibitem [{\citenamefont {Gudnason}\ \emph {et~al.}(2024)\citenamefont
  {Gudnason}, \citenamefont {Amari},\ and\ \citenamefont
  {Nitta}}]{Gudnason:2024shv}%
  \BibitemOpen
  \bibfield  {author} {\bibinfo {author} {\bibfnamefont {S.~B.}\ \bibnamefont
  {Gudnason}}, \bibinfo {author} {\bibfnamefont {Y.}~\bibnamefont {Amari}},\
  and\ \bibinfo {author} {\bibfnamefont {M.}~\bibnamefont {Nitta}},\ }\bibfield
   {title} {\bibinfo {title} {{Manipulation and creation of domain-wall
  skyrmions in chiral magnets}},\ }\href@noop {} {\  (\bibinfo {year}
  {2024})},\ \Eprint {https://arxiv.org/abs/2406.19056} {arXiv:2406.19056
  [cond-mat.mes-hall]} \BibitemShut {NoStop}%
\bibitem [{\citenamefont {Nagase}\ \emph {et~al.}(2021)\citenamefont {Nagase},
  \citenamefont {So}, \citenamefont {Yasui}, \citenamefont {Ishida},
  \citenamefont {Yoshida}, \citenamefont {Tanaka}, \citenamefont {Saitoh},
  \citenamefont {Ikarashi}, \citenamefont {Kawaguchi}, \citenamefont
  {Kuwahara},\ and\ \citenamefont {Nagao}}]{Nagase:2020imn}%
  \BibitemOpen
  \bibfield  {author} {\bibinfo {author} {\bibfnamefont {T.}~\bibnamefont
  {Nagase}}, \bibinfo {author} {\bibfnamefont {Y.-G.}\ \bibnamefont {So}},
  \bibinfo {author} {\bibfnamefont {H.}~\bibnamefont {Yasui}}, \bibinfo
  {author} {\bibfnamefont {T.}~\bibnamefont {Ishida}}, \bibinfo {author}
  {\bibfnamefont {H.~K.}\ \bibnamefont {Yoshida}}, \bibinfo {author}
  {\bibfnamefont {Y.}~\bibnamefont {Tanaka}}, \bibinfo {author} {\bibfnamefont
  {K.}~\bibnamefont {Saitoh}}, \bibinfo {author} {\bibfnamefont
  {N.}~\bibnamefont {Ikarashi}}, \bibinfo {author} {\bibfnamefont
  {Y.}~\bibnamefont {Kawaguchi}}, \bibinfo {author} {\bibfnamefont
  {M.}~\bibnamefont {Kuwahara}},\ and\ \bibinfo {author} {\bibfnamefont
  {M.}~\bibnamefont {Nagao}},\ }\bibfield  {title} {\bibinfo {title}
  {{Observation of domain wall bimerons in chiral magnets}},\ }\href
  {https://doi.org/10.1038/s41467-021-23845-y} {\bibfield  {journal} {\bibinfo
  {journal} {Nat. Commun.}\ }\textbf {\bibinfo {volume} {12}},\ \bibinfo
  {pages} {3490} (\bibinfo {year} {2021})}\BibitemShut {NoStop}%
\bibitem [{\citenamefont {Li}\ \emph {et~al.}(2021)\citenamefont {Li},
  \citenamefont {Rai}, \citenamefont {Pokhrel}, \citenamefont {Sapkota},
  \citenamefont {Mewes}, \citenamefont {Mewes}, \citenamefont {Xiao},
  \citenamefont {De~Graef},\ and\ \citenamefont {Sokalski}}]{li2021magnetic}%
  \BibitemOpen
  \bibfield  {author} {\bibinfo {author} {\bibfnamefont {M.}~\bibnamefont
  {Li}}, \bibinfo {author} {\bibfnamefont {A.}~\bibnamefont {Rai}}, \bibinfo
  {author} {\bibfnamefont {A.}~\bibnamefont {Pokhrel}}, \bibinfo {author}
  {\bibfnamefont {A.}~\bibnamefont {Sapkota}}, \bibinfo {author} {\bibfnamefont
  {C.}~\bibnamefont {Mewes}}, \bibinfo {author} {\bibfnamefont
  {T.}~\bibnamefont {Mewes}}, \bibinfo {author} {\bibfnamefont
  {D.}~\bibnamefont {Xiao}}, \bibinfo {author} {\bibfnamefont {M.}~\bibnamefont
  {De~Graef}},\ and\ \bibinfo {author} {\bibfnamefont {V.}~\bibnamefont
  {Sokalski}},\ }\bibfield  {title} {\bibinfo {title} {{Magnetic domain wall
  substructures in Pt/Co/Ni/Ir multi-layers}},\ }\href
  {https://doi.org/10.1063/5.0056100} {\bibfield  {journal} {\bibinfo
  {journal} {J. Appl. Phys.}\ }\textbf {\bibinfo {volume} {130}},\ \bibinfo
  {pages} {153903} (\bibinfo {year} {2021})}\BibitemShut {NoStop}%
\bibitem [{\citenamefont {Yang}\ \emph {et~al.}(2021)\citenamefont {Yang},
  \citenamefont {Nagase},\ and\ \citenamefont {Hirayama~et.al.}}]{Yang:2021}%
  \BibitemOpen
  \bibfield  {author} {\bibinfo {author} {\bibfnamefont {K.}~\bibnamefont
  {Yang}}, \bibinfo {author} {\bibfnamefont {K.}~\bibnamefont {Nagase}},\ and\
  \bibinfo {author} {\bibfnamefont {Y.}~\bibnamefont {Hirayama~et.al.}},\
  }\bibfield  {title} {\bibinfo {title} {{Wigner solids of domain wall
  skyrmions}},\ }\href {https://doi.org/10.1038/s41467-021-26306-8} {\bibfield
  {journal} {\bibinfo  {journal} {Nat. Commun.}\ }\textbf {\bibinfo {volume}
  {12}},\ \bibinfo {pages} {6006} (\bibinfo {year} {2021})}\BibitemShut
  {NoStop}%
\bibitem [{\citenamefont {Skomski}(2008)}]{Skomski2008}%
  \BibitemOpen
  \bibfield  {author} {\bibinfo {author} {\bibfnamefont {R.}~\bibnamefont
  {Skomski}},\ }\href@noop {} {\emph {\bibinfo {title} {Simple {{Models}} of
  {{Magnetism}}}}}\ (\bibinfo  {publisher} {Oxford University Press},\ \bibinfo
  {year} {2008})\BibitemShut {NoStop}%
\bibitem [{\citenamefont {Derrick}(1964)}]{Derrick1964}%
  \BibitemOpen
  \bibfield  {author} {\bibinfo {author} {\bibfnamefont {G.~H.}\ \bibnamefont
  {Derrick}},\ }\bibfield  {title} {\bibinfo {title} {Comments on {{Nonlinear
  Wave Equations}} as {{Models}} for {{Elementary Particles}}},\ }\href
  {https://doi.org/10.1063/1.1704233} {\bibfield  {journal} {\bibinfo
  {journal} {J. Math. Phys.}\ }\textbf {\bibinfo {volume} {5}},\ \bibinfo
  {pages} {1252} (\bibinfo {year} {1964})}\BibitemShut {NoStop}%
\bibitem [{\citenamefont {Manton}(2009)}]{Manton2009}%
  \BibitemOpen
  \bibfield  {author} {\bibinfo {author} {\bibfnamefont {N.~S.}\ \bibnamefont
  {Manton}},\ }\bibfield  {title} {\bibinfo {title} {Scaling identities for
  solitons beyond {{Derrick}}'s theorem},\ }\href
  {https://doi.org/10.1063/1.3089582} {\bibfield  {journal} {\bibinfo
  {journal} {J. Math. Phys.}\ }\textbf {\bibinfo {volume} {50}},\ \bibinfo
  {pages} {032901} (\bibinfo {year} {2009})}\BibitemShut {NoStop}%
\bibitem [{\citenamefont {Voinescu}\ \emph {et~al.}(2020)\citenamefont
  {Voinescu}, \citenamefont {Tai},\ and\ \citenamefont
  {Smalyukh}}]{PhysRevLett.125.057201}%
  \BibitemOpen
  \bibfield  {author} {\bibinfo {author} {\bibfnamefont {R.}~\bibnamefont
  {Voinescu}}, \bibinfo {author} {\bibfnamefont {J.-S.~B.}\ \bibnamefont
  {Tai}},\ and\ \bibinfo {author} {\bibfnamefont {I.~I.}\ \bibnamefont
  {Smalyukh}},\ }\bibfield  {title} {\bibinfo {title} {Hopf solitons in helical
  and conical backgrounds of chiral magnetic solids},\ }\href
  {https://doi.org/10.1103/PhysRevLett.125.057201} {\bibfield  {journal}
  {\bibinfo  {journal} {Phys. Rev. Lett.}\ }\textbf {\bibinfo {volume} {125}},\
  \bibinfo {pages} {057201} (\bibinfo {year} {2020})}\BibitemShut {NoStop}%
\bibitem [{\citenamefont {Prei{\ss}inger}\ \emph {et~al.}(2021)\citenamefont
  {Prei{\ss}inger}, \citenamefont {Karube}, \citenamefont {Ehlers},
  \citenamefont {Szigeti}, \citenamefont {Krug~von Nidda}, \citenamefont
  {White}, \citenamefont {Ukleev}, \citenamefont {R{\o}nnow}, \citenamefont
  {Tokunaga}, \citenamefont {Kikkawa}, \citenamefont {Tokura}, \citenamefont
  {Taguchi},\ and\ \citenamefont {K{\'e}zsm{\'a}rki}}]{preissinger_npj_2021}%
  \BibitemOpen
  \bibfield  {author} {\bibinfo {author} {\bibfnamefont {M.}~\bibnamefont
  {Prei{\ss}inger}}, \bibinfo {author} {\bibfnamefont {K.}~\bibnamefont
  {Karube}}, \bibinfo {author} {\bibfnamefont {D.}~\bibnamefont {Ehlers}},
  \bibinfo {author} {\bibfnamefont {B.}~\bibnamefont {Szigeti}}, \bibinfo
  {author} {\bibfnamefont {H.-A.}\ \bibnamefont {Krug~von Nidda}}, \bibinfo
  {author} {\bibfnamefont {J.~S.}\ \bibnamefont {White}}, \bibinfo {author}
  {\bibfnamefont {V.}~\bibnamefont {Ukleev}}, \bibinfo {author} {\bibfnamefont
  {H.~M.}\ \bibnamefont {R{\o}nnow}}, \bibinfo {author} {\bibfnamefont
  {Y.}~\bibnamefont {Tokunaga}}, \bibinfo {author} {\bibfnamefont
  {A.}~\bibnamefont {Kikkawa}}, \bibinfo {author} {\bibfnamefont
  {Y.}~\bibnamefont {Tokura}}, \bibinfo {author} {\bibfnamefont
  {Y.}~\bibnamefont {Taguchi}},\ and\ \bibinfo {author} {\bibfnamefont
  {I.}~\bibnamefont {K{\'e}zsm{\'a}rki}},\ }\bibfield  {title} {\bibinfo
  {title} {Vital role of magnetocrystalline anisotropy in cubic chiral skyrmion
  hosts},\ }\href {https://doi.org/10.1038/s41535-021-00365-y} {\bibfield
  {journal} {\bibinfo  {journal} {npj Quantum Materials}\ }\textbf {\bibinfo
  {volume} {6}},\ \bibinfo {pages} {65} (\bibinfo {year} {2021})}\BibitemShut
  {NoStop}%
\bibitem [{\citenamefont {Wilson}\ \emph {et~al.}(2014)\citenamefont {Wilson},
  \citenamefont {Butenko}, \citenamefont {Bogdanov},\ and\ \citenamefont
  {Monchesky}}]{wilson_prb_2014}%
  \BibitemOpen
  \bibfield  {author} {\bibinfo {author} {\bibfnamefont {M.~N.}\ \bibnamefont
  {Wilson}}, \bibinfo {author} {\bibfnamefont {A.~B.}\ \bibnamefont {Butenko}},
  \bibinfo {author} {\bibfnamefont {A.~N.}\ \bibnamefont {Bogdanov}},\ and\
  \bibinfo {author} {\bibfnamefont {T.~L.}\ \bibnamefont {Monchesky}},\
  }\bibfield  {title} {\bibinfo {title} {Chiral skyrmions in cubic helimagnet
  films: The role of uniaxial anisotropy},\ }\href
  {https://doi.org/10.1103/PhysRevB.89.094411} {\bibfield  {journal} {\bibinfo
  {journal} {Phys. Rev. B}\ }\textbf {\bibinfo {volume} {89}},\ \bibinfo
  {pages} {094411} (\bibinfo {year} {2014})}\BibitemShut {NoStop}%
\bibitem [{\citenamefont {Leask}\ and\ \citenamefont
  {Speight}(2025)}]{leask_2025}%
  \BibitemOpen
  \bibfield  {author} {\bibinfo {author} {\bibfnamefont {P.}~\bibnamefont
  {Leask}}\ and\ \bibinfo {author} {\bibfnamefont {M.}~\bibnamefont
  {Speight}},\ }\href {https://arxiv.org/abs/2504.17772} {\bibinfo {title}
  {Demagnetization in micromagnetics: magnetostatic self-interactions of bulk
  chiral magnetic skyrmions}} (\bibinfo {year} {2025}),\ \Eprint
  {https://arxiv.org/abs/2504.17772} {arXiv:2504.17772 [cond-mat.mes-hall]}
  \BibitemShut {NoStop}%
\bibitem [{\citenamefont {Yu}\ \emph {et~al.}(2011)\citenamefont {Yu},
  \citenamefont {Kanazawa}, \citenamefont {Onose}, \citenamefont {Kimoto},
  \citenamefont {Zhang}, \citenamefont {Ishiwata}, \citenamefont {Matsui},\
  and\ \citenamefont {Tokura}}]{yu_natmat_2011}%
  \BibitemOpen
  \bibfield  {author} {\bibinfo {author} {\bibfnamefont {X.~Z.}\ \bibnamefont
  {Yu}}, \bibinfo {author} {\bibfnamefont {N.}~\bibnamefont {Kanazawa}},
  \bibinfo {author} {\bibfnamefont {Y.}~\bibnamefont {Onose}}, \bibinfo
  {author} {\bibfnamefont {K.}~\bibnamefont {Kimoto}}, \bibinfo {author}
  {\bibfnamefont {W.~Z.}\ \bibnamefont {Zhang}}, \bibinfo {author}
  {\bibfnamefont {S.}~\bibnamefont {Ishiwata}}, \bibinfo {author}
  {\bibfnamefont {Y.}~\bibnamefont {Matsui}},\ and\ \bibinfo {author}
  {\bibfnamefont {Y.}~\bibnamefont {Tokura}},\ }\bibfield  {title} {\bibinfo
  {title} {Near room-temperature formation of a skyrmion crystal in thin-films
  of the helimagnet {FeGe}},\ }\href {https://doi.org/10.1038/nmat2916}
  {\bibfield  {journal} {\bibinfo  {journal} {Nat. Mater.}\ }\textbf {\bibinfo
  {volume} {10}},\ \bibinfo {pages} {106} (\bibinfo {year} {2011})}\BibitemShut
  {NoStop}%
\bibitem [{\citenamefont {Yu}\ \emph {et~al.}(2018)\citenamefont {Yu},
  \citenamefont {Koshibae}, \citenamefont {Tokunaga}, \citenamefont {Shibata},
  \citenamefont {Taguchi}, \citenamefont {Nagaosa},\ and\ \citenamefont
  {Tokura}}]{yu_nature_2018}%
  \BibitemOpen
  \bibfield  {author} {\bibinfo {author} {\bibfnamefont {X.~Z.}\ \bibnamefont
  {Yu}}, \bibinfo {author} {\bibfnamefont {W.}~\bibnamefont {Koshibae}},
  \bibinfo {author} {\bibfnamefont {Y.}~\bibnamefont {Tokunaga}}, \bibinfo
  {author} {\bibfnamefont {K.}~\bibnamefont {Shibata}}, \bibinfo {author}
  {\bibfnamefont {Y.}~\bibnamefont {Taguchi}}, \bibinfo {author} {\bibfnamefont
  {N.}~\bibnamefont {Nagaosa}},\ and\ \bibinfo {author} {\bibfnamefont
  {Y.}~\bibnamefont {Tokura}},\ }\bibfield  {title} {\bibinfo {title}
  {Transformation between meron and skyrmion topological spin textures in a
  chiral magnet},\ }\href {https://doi.org/10.1038/s41586-018-0745-3}
  {\bibfield  {journal} {\bibinfo  {journal} {Nature}\ }\textbf {\bibinfo
  {volume} {564}},\ \bibinfo {pages} {95} (\bibinfo {year} {2018})}\BibitemShut
  {NoStop}%
\bibitem [{\citenamefont {Ukleev}\ \emph {et~al.}(2024)\citenamefont {Ukleev},
  \citenamefont {Utesov}, \citenamefont {Luo}, \citenamefont {Radu},
  \citenamefont {Wintz}, \citenamefont {Weigand}, \citenamefont {Finizio},
  \citenamefont {Winter}, \citenamefont {Tahn}, \citenamefont {Rellinghaus},
  \citenamefont {Karube}, \citenamefont {Tokura}, \citenamefont {Taguchi},\
  and\ \citenamefont {White}}]{ukleev_prb_2024}%
  \BibitemOpen
  \bibfield  {author} {\bibinfo {author} {\bibfnamefont {V.}~\bibnamefont
  {Ukleev}}, \bibinfo {author} {\bibfnamefont {O.~I.}\ \bibnamefont {Utesov}},
  \bibinfo {author} {\bibfnamefont {C.}~\bibnamefont {Luo}}, \bibinfo {author}
  {\bibfnamefont {F.}~\bibnamefont {Radu}}, \bibinfo {author} {\bibfnamefont
  {S.}~\bibnamefont {Wintz}}, \bibinfo {author} {\bibfnamefont
  {M.}~\bibnamefont {Weigand}}, \bibinfo {author} {\bibfnamefont
  {S.}~\bibnamefont {Finizio}}, \bibinfo {author} {\bibfnamefont
  {M.}~\bibnamefont {Winter}}, \bibinfo {author} {\bibfnamefont
  {A.}~\bibnamefont {Tahn}}, \bibinfo {author} {\bibfnamefont {B.}~\bibnamefont
  {Rellinghaus}}, \bibinfo {author} {\bibfnamefont {K.}~\bibnamefont {Karube}},
  \bibinfo {author} {\bibfnamefont {Y.}~\bibnamefont {Tokura}}, \bibinfo
  {author} {\bibfnamefont {Y.}~\bibnamefont {Taguchi}},\ and\ \bibinfo {author}
  {\bibfnamefont {J.~S.}\ \bibnamefont {White}},\ }\bibfield  {title} {\bibinfo
  {title} {Competing anisotropies in the chiral cubic magnet
  {${\mathrm{Co}}_{8}{\mathrm{Zn}}_{8}{\mathrm{Mn}}_{4}$} unveiled by resonant
  x-ray magnetic scattering},\ }\href
  {https://doi.org/10.1103/PhysRevB.109.184415} {\bibfield  {journal} {\bibinfo
   {journal} {Phys. Rev. B}\ }\textbf {\bibinfo {volume} {109}},\ \bibinfo
  {pages} {184415} (\bibinfo {year} {2024})}\BibitemShut {NoStop}%
\bibitem [{\citenamefont {Lin}\ \emph {et~al.}(2015)\citenamefont {Lin},
  \citenamefont {Saxena},\ and\ \citenamefont {Batista}}]{Lin2015}%
  \BibitemOpen
  \bibfield  {author} {\bibinfo {author} {\bibfnamefont {S.-Z.}\ \bibnamefont
  {Lin}}, \bibinfo {author} {\bibfnamefont {A.}~\bibnamefont {Saxena}},\ and\
  \bibinfo {author} {\bibfnamefont {C.~D.}\ \bibnamefont {Batista}},\
  }\bibfield  {title} {\bibinfo {title} {Skyrmion fractionalization and merons
  in chiral magnets with easy-plane anisotropy},\ }\href
  {https://doi.org/10.1103/PhysRevB.91.224407} {\bibfield  {journal} {\bibinfo
  {journal} {Phys. Rev. B}\ }\textbf {\bibinfo {volume} {91}},\ \bibinfo
  {pages} {224407} (\bibinfo {year} {2015})}\BibitemShut {NoStop}%
\bibitem [{\citenamefont {Harland}\ \emph {et~al.}(2023)\citenamefont
  {Harland}, \citenamefont {Leask},\ and\ \citenamefont
  {Speight}}]{Harland_jmathphys_2023}%
  \BibitemOpen
  \bibfield  {author} {\bibinfo {author} {\bibfnamefont {D.}~\bibnamefont
  {Harland}}, \bibinfo {author} {\bibfnamefont {P.}~\bibnamefont {Leask}},\
  and\ \bibinfo {author} {\bibfnamefont {M.}~\bibnamefont {Speight}},\
  }\bibfield  {title} {\bibinfo {title} {{Skyrme crystals with massive
  pions}},\ }\href {https://doi.org/10.1063/5.0159674} {\bibfield  {journal}
  {\bibinfo  {journal} {J. Math. Phys.}\ }\textbf {\bibinfo {volume} {64}},\
  \bibinfo {pages} {103503} (\bibinfo {year} {2023})}\BibitemShut {NoStop}%
\bibitem [{\citenamefont {Leask}\ \emph {et~al.}(2024)\citenamefont {Leask},
  \citenamefont {Huidobro},\ and\ \citenamefont {Wereszczynski}}]{Leask2024a}%
  \BibitemOpen
  \bibfield  {author} {\bibinfo {author} {\bibfnamefont {P.}~\bibnamefont
  {Leask}}, \bibinfo {author} {\bibfnamefont {M.}~\bibnamefont {Huidobro}},\
  and\ \bibinfo {author} {\bibfnamefont {A.}~\bibnamefont {Wereszczynski}},\
  }\bibfield  {title} {\bibinfo {title} {Generalized skyrmion crystals with
  applications to neutron stars},\ }\href
  {https://doi.org/10.1103/PhysRevD.109.056013} {\bibfield  {journal} {\bibinfo
   {journal} {Phys. Rev. D}\ }\textbf {\bibinfo {volume} {109}},\ \bibinfo
  {pages} {056013} (\bibinfo {year} {2024})}\BibitemShut {NoStop}%
\bibitem [{\citenamefont {Harland}\ \emph {et~al.}(2024)\citenamefont
  {Harland}, \citenamefont {Leask},\ and\ \citenamefont
  {Speight}}]{Harland:2024}%
  \BibitemOpen
  \bibfield  {author} {\bibinfo {author} {\bibfnamefont {D.}~\bibnamefont
  {Harland}}, \bibinfo {author} {\bibfnamefont {P.}~\bibnamefont {Leask}},\
  and\ \bibinfo {author} {\bibfnamefont {M.}~\bibnamefont {Speight}},\
  }\bibfield  {title} {\bibinfo {title} {Skyrmion crystals stabilized by
  $\omega$-mesons},\ }\href {https://doi.org/10.1007/JHEP06(2024)116}
  {\bibfield  {journal} {\bibinfo  {journal} {J. High Energy Phys.}\ }\textbf
  {\bibinfo {volume} {06}}\bibinfo  {number} { (6)},\ \bibinfo {pages}
  {116}}\BibitemShut {NoStop}%
\bibitem [{\citenamefont {Speight}\ and\ \citenamefont
  {Winyard}(2024)}]{speight2024}%
  \BibitemOpen
\bibfield  {number} {  }\bibfield  {author} {\bibinfo {author} {\bibfnamefont
  {M.}~\bibnamefont {Speight}}\ and\ \bibinfo {author} {\bibfnamefont
  {T.}~\bibnamefont {Winyard}},\ }\href {https://arxiv.org/abs/2406.16584}
  {\bibinfo {title} {Vortex lattices and critical fields in anisotropic
  superconductors}} (\bibinfo {year} {2024}),\ \Eprint
  {https://arxiv.org/abs/2406.16584} {arXiv:2406.16584 [cond-mat.supr-con]}
  \BibitemShut {NoStop}%
\bibitem [{\citenamefont {Speight}(2014)}]{Speight2014}%
  \BibitemOpen
  \bibfield  {author} {\bibinfo {author} {\bibfnamefont {J.~M.}\ \bibnamefont
  {Speight}},\ }\bibfield  {title} {\bibinfo {title} {Solitons on {{Tori}} and
  {{Soliton Crystals}}},\ }\href {https://doi.org/10.1007/s00220-014-2104-z}
  {\bibfield  {journal} {\bibinfo  {journal} {Commun. Math. Phys.}\ }\textbf
  {\bibinfo {volume} {332}},\ \bibinfo {pages} {355} (\bibinfo {year}
  {2014})}\BibitemShut {NoStop}%
\bibitem [{\citenamefont {Gudnason}\ and\ \citenamefont
  {Speight}(2020)}]{gudnason_jhep_2020}%
  \BibitemOpen
  \bibfield  {author} {\bibinfo {author} {\bibfnamefont {S.~B.}\ \bibnamefont
  {Gudnason}}\ and\ \bibinfo {author} {\bibfnamefont {J.~M.}\ \bibnamefont
  {Speight}},\ }\bibfield  {title} {\bibinfo {title} {Realistic classical
  binding energies in the $\omega$-skyrme model},\ }\href
  {https://doi.org/10.1007/JHEP07(2020)184} {\bibfield  {journal} {\bibinfo
  {journal} {J. High Energy Phys.}\ }\textbf {\bibinfo {volume} {07}}\bibinfo
  {number} { (7)},\ \bibinfo {pages} {184}}\BibitemShut {NoStop}%
\bibitem [{\citenamefont {Gudnason}\ and\ \citenamefont
  {Halcrow}(2022)}]{gudnason_jhep_2022}%
  \BibitemOpen
\bibfield  {number} {  }\bibfield  {author} {\bibinfo {author} {\bibfnamefont
  {S.~B.}\ \bibnamefont {Gudnason}}\ and\ \bibinfo {author} {\bibfnamefont
  {C.}~\bibnamefont {Halcrow}},\ }\bibfield  {title} {\bibinfo {title} {A
  sm{\"o}rg{\aa}sbord of skyrmions},\ }\href
  {https://doi.org/10.1007/JHEP08(2022)117} {\bibfield  {journal} {\bibinfo
  {journal} {J. High Energy Phys.}\ }\textbf {\bibinfo {volume} {08}}\bibinfo
  {number} { (8)},\ \bibinfo {pages} {117}}\BibitemShut {NoStop}%
\bibitem [{\citenamefont {Leask}(2022)}]{Leask2022}%
  \BibitemOpen
\bibfield  {number} {  }\bibfield  {author} {\bibinfo {author} {\bibfnamefont
  {P.}~\bibnamefont {Leask}},\ }\bibfield  {title} {\bibinfo {title} {Baby
  {{Skyrmion}} crystals},\ }\href {https://doi.org/10.1103/PhysRevD.105.025010}
  {\bibfield  {journal} {\bibinfo  {journal} {Phys. Rev. D}\ }\textbf {\bibinfo
  {volume} {105}},\ \bibinfo {pages} {025010} (\bibinfo {year}
  {2022})}\BibitemShut {NoStop}%
\bibitem [{Note1()}]{Note1}%
  \BibitemOpen
  \bibinfo {note} {The Supplemental Material includes the details of the
  metric-optimization method.}\BibitemShut {Stop}%
\bibitem [{\citenamefont {Schroers}(2019)}]{schroers_scipost_2019}%
  \BibitemOpen
  \bibfield  {author} {\bibinfo {author} {\bibfnamefont {B.~J.}\ \bibnamefont
  {Schroers}},\ }\bibfield  {title} {\bibinfo {title} {{Gauged sigma models and
  magnetic Skyrmions}},\ }\href {https://doi.org/10.21468/SciPostPhys.7.3.030}
  {\bibfield  {journal} {\bibinfo  {journal} {SciPost Phys.}\ }\textbf
  {\bibinfo {volume} {7}},\ \bibinfo {pages} {030} (\bibinfo {year}
  {2019})}\BibitemShut {NoStop}%
\bibitem [{\citenamefont {Hill}\ \emph {et~al.}(2021)\citenamefont {Hill},
  \citenamefont {Slastikov},\ and\ \citenamefont
  {Tchernyshyov}}]{hill_scipost_2021}%
  \BibitemOpen
  \bibfield  {author} {\bibinfo {author} {\bibfnamefont {D.}~\bibnamefont
  {Hill}}, \bibinfo {author} {\bibfnamefont {V.}~\bibnamefont {Slastikov}},\
  and\ \bibinfo {author} {\bibfnamefont {O.}~\bibnamefont {Tchernyshyov}},\
  }\bibfield  {title} {\bibinfo {title} {{Chiral magnetism: a geometric
  perspective}},\ }\href {https://doi.org/10.21468/SciPostPhys.10.3.078}
  {\bibfield  {journal} {\bibinfo  {journal} {SciPost Phys.}\ }\textbf
  {\bibinfo {volume} {10}},\ \bibinfo {pages} {078} (\bibinfo {year}
  {2021})}\BibitemShut {NoStop}%
\bibitem [{\citenamefont {Amari}\ \emph
  {et~al.}(2024{\natexlab{d}})\citenamefont {Amari}, \citenamefont {Eto},\ and\
  \citenamefont {Nitta}}]{amari_jhep_2024}%
  \BibitemOpen
  \bibfield  {author} {\bibinfo {author} {\bibfnamefont {Y.}~\bibnamefont
  {Amari}}, \bibinfo {author} {\bibfnamefont {M.}~\bibnamefont {Eto}},\ and\
  \bibinfo {author} {\bibfnamefont {M.}~\bibnamefont {Nitta}},\ }\bibfield
  {title} {\bibinfo {title} {Topological solitons stabilized by a background
  gauge field and soliton-anti-soliton asymmetry},\ }\href
  {https://doi.org/10.1007/JHEP11(2024)127} {\bibfield  {journal} {\bibinfo
  {journal} {J. High Energy Phys.}\ }\textbf {\bibinfo {volume} {2024}}\bibinfo
   {number} { (11)},\ \bibinfo {pages} {127}}\BibitemShut {NoStop}%
\bibitem [{\citenamefont {Garcia-Sanchez}\ \emph {et~al.}(2015)\citenamefont
  {Garcia-Sanchez}, \citenamefont {Borys}, \citenamefont {Soucaille},
  \citenamefont {Adam}, \citenamefont {Stamps},\ and\ \citenamefont
  {Kim}}]{garcia-sanchez_prl_2015}%
  \BibitemOpen
\bibfield  {number} {  }\bibfield  {author} {\bibinfo {author} {\bibfnamefont
  {F.}~\bibnamefont {Garcia-Sanchez}}, \bibinfo {author} {\bibfnamefont
  {P.}~\bibnamefont {Borys}}, \bibinfo {author} {\bibfnamefont
  {R.}~\bibnamefont {Soucaille}}, \bibinfo {author} {\bibfnamefont {J.-P.}\
  \bibnamefont {Adam}}, \bibinfo {author} {\bibfnamefont {R.~L.}\ \bibnamefont
  {Stamps}},\ and\ \bibinfo {author} {\bibfnamefont {J.-V.}\ \bibnamefont
  {Kim}},\ }\bibfield  {title} {\bibinfo {title} {Narrow magnonic waveguides
  based on domain walls},\ }\href
  {https://doi.org/10.1103/PhysRevLett.114.247206} {\bibfield  {journal}
  {\bibinfo  {journal} {Phys. Rev. Lett.}\ }\textbf {\bibinfo {volume} {114}},\
  \bibinfo {pages} {247206} (\bibinfo {year} {2015})}\BibitemShut {NoStop}%
\bibitem [{\citenamefont {Lan}\ \emph {et~al.}(2015)\citenamefont {Lan},
  \citenamefont {Yu}, \citenamefont {Wu},\ and\ \citenamefont
  {Xiao}}]{lan_prx_2015}%
  \BibitemOpen
  \bibfield  {author} {\bibinfo {author} {\bibfnamefont {J.}~\bibnamefont
  {Lan}}, \bibinfo {author} {\bibfnamefont {W.}~\bibnamefont {Yu}}, \bibinfo
  {author} {\bibfnamefont {R.}~\bibnamefont {Wu}},\ and\ \bibinfo {author}
  {\bibfnamefont {J.}~\bibnamefont {Xiao}},\ }\bibfield  {title} {\bibinfo
  {title} {Spin-wave diode},\ }\href
  {https://doi.org/10.1103/PhysRevX.5.041049} {\bibfield  {journal} {\bibinfo
  {journal} {Phys. Rev. X}\ }\textbf {\bibinfo {volume} {5}},\ \bibinfo {pages}
  {041049} (\bibinfo {year} {2015})}\BibitemShut {NoStop}%
\bibitem [{\citenamefont {Wagner}\ \emph {et~al.}(2016)\citenamefont {Wagner},
  \citenamefont {K{\'a}kay}, \citenamefont {Schultheiss}, \citenamefont
  {Henschke}, \citenamefont {Sebastian},\ and\ \citenamefont
  {Schultheiss}}]{wagner_natnano_2016}%
  \BibitemOpen
  \bibfield  {author} {\bibinfo {author} {\bibfnamefont {K.}~\bibnamefont
  {Wagner}}, \bibinfo {author} {\bibfnamefont {A.}~\bibnamefont {K{\'a}kay}},
  \bibinfo {author} {\bibfnamefont {K.}~\bibnamefont {Schultheiss}}, \bibinfo
  {author} {\bibfnamefont {A.}~\bibnamefont {Henschke}}, \bibinfo {author}
  {\bibfnamefont {T.}~\bibnamefont {Sebastian}},\ and\ \bibinfo {author}
  {\bibfnamefont {H.}~\bibnamefont {Schultheiss}},\ }\bibfield  {title}
  {\bibinfo {title} {Magnetic domain walls as reconfigurable spin-wave
  nanochannels},\ }\href {https://doi.org/10.1038/nnano.2015.339} {\bibfield
  {journal} {\bibinfo  {journal} {Nature Nanotechnology}\ }\textbf {\bibinfo
  {volume} {11}},\ \bibinfo {pages} {432} (\bibinfo {year} {2016})}\BibitemShut
  {NoStop}%
\bibitem [{\citenamefont {Sluka}\ \emph {et~al.}(2019)\citenamefont {Sluka},
  \citenamefont {Schneider}, \citenamefont {Gallardo}, \citenamefont
  {K{\'a}kay}, \citenamefont {Weigand}, \citenamefont {Warnatz}, \citenamefont
  {Mattheis}, \citenamefont {Rold{\'a}n-Molina}, \citenamefont {Landeros},
  \citenamefont {Tiberkevich}, \citenamefont {Slavin}, \citenamefont
  {Sch{\"u}tz}, \citenamefont {Erbe}, \citenamefont {Deac}, \citenamefont
  {Lindner}, \citenamefont {Raabe}, \citenamefont {Fassbender},\ and\
  \citenamefont {Wintz}}]{sluka_natnano_2019}%
  \BibitemOpen
  \bibfield  {author} {\bibinfo {author} {\bibfnamefont {V.}~\bibnamefont
  {Sluka}}, \bibinfo {author} {\bibfnamefont {T.}~\bibnamefont {Schneider}},
  \bibinfo {author} {\bibfnamefont {R.~A.}\ \bibnamefont {Gallardo}}, \bibinfo
  {author} {\bibfnamefont {A.}~\bibnamefont {K{\'a}kay}}, \bibinfo {author}
  {\bibfnamefont {M.}~\bibnamefont {Weigand}}, \bibinfo {author} {\bibfnamefont
  {T.}~\bibnamefont {Warnatz}}, \bibinfo {author} {\bibfnamefont
  {R.}~\bibnamefont {Mattheis}}, \bibinfo {author} {\bibfnamefont
  {A.}~\bibnamefont {Rold{\'a}n-Molina}}, \bibinfo {author} {\bibfnamefont
  {P.}~\bibnamefont {Landeros}}, \bibinfo {author} {\bibfnamefont
  {V.}~\bibnamefont {Tiberkevich}}, \bibinfo {author} {\bibfnamefont
  {A.}~\bibnamefont {Slavin}}, \bibinfo {author} {\bibfnamefont
  {G.}~\bibnamefont {Sch{\"u}tz}}, \bibinfo {author} {\bibfnamefont
  {A.}~\bibnamefont {Erbe}}, \bibinfo {author} {\bibfnamefont {A.}~\bibnamefont
  {Deac}}, \bibinfo {author} {\bibfnamefont {J.}~\bibnamefont {Lindner}},
  \bibinfo {author} {\bibfnamefont {J.}~\bibnamefont {Raabe}}, \bibinfo
  {author} {\bibfnamefont {J.}~\bibnamefont {Fassbender}},\ and\ \bibinfo
  {author} {\bibfnamefont {S.}~\bibnamefont {Wintz}},\ }\bibfield  {title}
  {\bibinfo {title} {Emission and propagation of 1d and 2d spin waves with
  nanoscale wavelengths in anisotropic spin textures},\ }\href
  {https://doi.org/10.1038/s41565-019-0383-4} {\bibfield  {journal} {\bibinfo
  {journal} {Nature Nanotechnology}\ }\textbf {\bibinfo {volume} {14}},\
  \bibinfo {pages} {328} (\bibinfo {year} {2019})}\BibitemShut {NoStop}%
\end{thebibliography}%

\clearpage
\newpage
\widetext
\begin{center}
    \textbf{\large Supplemental material for ``Domain Wall Networks as Skyrmion Crystals in Chiral Magnets''}\\
    \vspace{0.5cm}
    \text{Seungho Lee, Toshiaki Fujimori, Muneto Nitta, and Se Kwon Kim}
\end{center}

\renewcommand{\theequation}{S\arabic{equation}}
\renewcommand{\thefigure}{S\arabic{figure}}
\setcounter{equation}{0}
\setcounter{figure}{0}
\setcounter{page}{1}

The Supplemental Material includes the details of the metric-optimization method.

\section{Metric optimization}\label{supp_metric}
The energy for the chiral magnet contains the Dirichlet, Dzyaloshinskii-Moriya, and potential terms, which are given by
\begin{align}
    E &= E_2 + E_1 + E_0\,, \label{energy}
    \\
    E_2 &= \int_{\mathbb{R}^2/\Lambda} d^2 x \left[\frac{1}{2} \sum_i \partial_i \mathbf{n}\cdot \partial_i \mathbf{n}\right]\,,
    \\
    E_1 &= \int_{\mathbb{R}^2/\Lambda} d^2 x D \mathbf{n}\cdot \left(\bm\nabla\times\mathbf{n}\right) \nonumber
    \\
    &= \int_{\mathbb{R}^2/\Lambda} d^2 x 2D n_3\left(\partial_1 n_2 - \partial_2 n_1\right) \,,
    \\
    E_0 &= \int_{\mathbb{R}^2/\Lambda} d^2 x \mathcal{E}_0 (\mathbf{n})\,.
\end{align}
Here, the energy functional is defined on the unit cell $\mathbb{R}^2/\Lambda$ with the lattice
\begin{equation}
    \Lambda = \{ n_1 \mathbf{v}_1 + n_2 \mathbf{v}_2\, | \,n_i \in \mathbb{Z} , \mathbf{v}_i \in \mathbb{R}^2 \}\,,
\end{equation}
where $\mathbf{v}_1$ and $\mathbf{v}_2$ are primitive vectors. The geometry of the unit cell is determined by the primitive vectors or, equivalently, the matrix $L = (\mathbf{v}_1\,\,\mathbf{v}_2)$. The size of the unit cell is determined by the area $\mathcal{A} = \det L$, and the shape is determined by the angle and the relative length between the primitive vectors. The unit cell can be mapped from the torus by the parametrization
\begin{align}
    \mathbf{x} = X_1 \mathbf{v}_1 + X_2 \mathbf{v}_2\,,
\end{align}
where $\mathbf{x} = (x_1, x_2) \in \mathbb{R}^2/\Lambda$ and $(X_1, X_2) \in [0,1]^2$. The variables $(x_1, x_2)$ and $(X_1, X_2)$ can be transformed by the relation $\partial X_i/ \partial x_j = (L^{-1})_{ij}$ from the definition $X_i = (L^{-1})_{ij}x_j$. To obtain the ground state solution, we find the optimal set of the field $\mathbf{n}$ and the matrix $L$, which minimize the energy density. To this end, we first express the energy as a function of $L$. With $(X_1, X_2)$ and $L$, the energy can be written as
\begin{align}
    E_2 &= \int_{\mathbb{R}^2/\Lambda} d^2 x \frac{1}{2} \sum_i \partial_i \mathbf{n} \cdot \partial_i \mathbf{n} \nonumber
    \\
    &=\int_{[0,1]^2} d^2 X \det L \frac{1}{2} \sum_i \left(\sum_j \frac{\partial \mathbf{n}}{\partial X_j} \frac{\partial X_j}{\partial x_i}\right) \cdot \left( \sum_k\frac{\partial \mathbf{n}}{\partial X_k} \frac{\partial X_k}{\partial x_i}\right) \nonumber
    \\
    &=\int_{[0,1]^2} d^2 X \det L \frac{1}{2} \sum_{ijk}  \frac{\partial \mathbf{n}}{\partial X_j} (L^{-1})_{ji} \cdot  \frac{\partial \mathbf{n}}{\partial X_k} (L^{-1})_{ki} \nonumber
    \\
    &=\int_{[0,1]^2} d^2 X \det{L} \frac{1}{2} \sum_{ijk}  \frac{\partial \mathbf{n}}{\partial X_j}  \cdot  \frac{\partial \mathbf{n}}{\partial X_k} (L^{-1})_{ji} (L^{-1})_{ki}  \nonumber
    \\
    &= (-1)\int_{[0,1]^2} d^2 X \det{L} \frac{1}{2} \sum_{ijk}  \mathbf{n}  \cdot  \frac{\partial^2 \mathbf{n}}{\partial X_j \partial X_k}(L^{-1})_{ji} (L^{-1})_{ki} \,,
\end{align}
\begin{align}
    E_1 &= \int_{\mathbb{R}^2/\Lambda} d^2 x D \mathbf{n}\cdot \left(\bm\nabla\times\mathbf{n}\right) \nonumber
    \\
    &= \int_{\mathbb{R}^2/\Lambda} d^2 x 2D n_3\left(\partial_1 n_2 - \partial_2 n_1\right)  \nonumber
    \\
    &= \int_{[0,1]^2} d^2 X \det{L} 2D n_3 \sum_i\left(\frac{\partial n_2}{\partial X_i} (L^{-1})_{i1} - \frac{\partial n_1}{\partial X_i} (L^{-1})_{i2} \right)\,,
    \\
    E_0 &= \int_{[0,1]^2} d^2 X \det L \mathcal{E}_0 (\mathbf{n})\,.
\end{align}
Then the energy density $E/
\mathcal{A}=E/\det L$ is
\begin{align}
    \frac{E}{\mathcal{A}} = \int_{[0,1]^2} d^2 X \left[\frac{1}{2}\sum_{ijk} \frac{\partial \mathbf{n}}{\partial X_j} \cdot \frac{\partial \mathbf{n}}{\partial X_k} M_{ji} M_{ki} + 2D n_3 \sum_i \left( \frac{\partial n_2}{\partial X_i}M_{i1} - \frac{\partial n_1}{\partial X_i}M_{i2}\right) + \mathcal{E}_0 (\mathbf{n})\right]\,,
\end{align}
where $M=L^{-1}$.
Also, we can express the topological charge on the torus as
\begin{align}
    Q &= \frac{1}{4\pi} \int_{[0,1]^2}  d^2X \mathbf{n} \cdot \left( \sum_{ij}\frac{\partial \mathbf{n}}{\partial X_i} \times \frac{\partial \mathbf{n}}{\partial X_j} M_{i1}M_{j2} \right) \det L \,.
\end{align}
To minimize $E/\mathcal{A}$, we use the arrested Newton flow algorithm, which solves the Newtonian equation of motion 
\begin{align}
    \ddot{n}_\alpha &= -\frac{\delta}{\delta n_\alpha}\left(\frac{E}{\mathcal{A}}\right)\,,
    \\
    \ddot{M}_{\alpha \beta} &= -\frac{\partial}{\partial M_{\alpha \beta}}\left(\frac{E}{\mathcal{A}}\right)\,,
\end{align}
with the arrest process that sets $\dot{\mathbf{n}}(t+\delta t)=0$ when $\ddot{\mathbf{n}}(t+\delta t)\cdot \ddot{\mathbf{n}}(t) < 0$ and sets $\dot{M}(t + \delta t) = 0$ when $\min_{ij}\ddot{M}_{ij}(t+\delta t) \ddot{M}_{ij} (t) < 0$. The gradient of the energy density with respect to the field is obtained by
\begin{align}
      \frac{\delta }{\delta n_\alpha}\left( \frac{E_2}{\mathcal{A}} \right) &= \sum_{ijk} (-1) \frac{\partial^2 n_\alpha}{\partial X_j \partial X_k} M_{ji}M_{ki}\,,
    \\
     \frac{\delta }{\delta n_1}\left( \frac{E_1}{\mathcal{A}} \right) &= 2D \sum_i \frac{\partial n_3}{\partial X_i}M_{i2}\,,
    \\
    \frac{\delta }{\delta n_2}\left( \frac{E_1}{\mathcal{A}} \right) &= -2D \sum_i \frac{\partial n_3}{\partial X_i}M_{i1}\,,
    \\
    \frac{\delta }{\delta n_3}\left( \frac{E_1}{\mathcal{A}} \right) &=  2D \sum_i\left(\frac{\partial n_2}{\partial X_i} M_{i1} - \frac{\partial n_1}{\partial X_i} M_{i2} \right)\,.
    \\
    \frac{\delta }{\delta n_\alpha}\left( \frac{E_0}{\mathcal{A}} \right) &= \frac{\partial \mathcal{E}_0}{\partial n_\alpha}\,.
\end{align}
Similarly, the gradient with respect to $M$ is
\begin{align}
    \frac{\partial }{\partial M_{\alpha\beta}}\left( \frac{E_2}{\mathcal{A}} \right)&= \int_{[0,1]^2} d^2X (-1)\sum_k \mathbf{n} \cdot \frac{\partial^2 \mathbf{n}}{\partial X_\alpha\partial X_k} M_{k\beta}\,,
    \\
    \frac{\partial}{\partial M_{\alpha\beta}} \left( \frac{E_1}{\mathcal{A}}\right) &= \int_{[0,1]^2} d^2 X 2D n_3 \left(\frac{\partial n_2}{\partial X_\alpha} \delta_{1\beta} - \frac{\partial n_1}{\partial X_\alpha} \delta_{2\beta} \right)\,,
    \\
    \frac{\partial}{\partial M_{\alpha\beta}} \left( \frac{E_0}{\mathcal{A}}\right) &= 0\,.
\end{align}
We iteratively apply this algorithm for $\mathbf{n}$ and $M$ until $\ddot{n}_\alpha = 0 $ and $ \ddot{M}_{\alpha \beta} = 0$.

\end{document}